\documentclass[12pt]{article}
\usepackage{epsf}
\usepackage{bm,amssymb}
\usepackage{hyperref}
\usepackage{graphicx}

\newtheorem{theorem}{Theorem}
\newtheorem{lemma}{Lemma}

\newcommand{\be}{{\bf E}}
\newcommand{\Tr}{{\rm Tr}}
\newcommand{\Sref}{Section~\ref}
\newcommand{\sref}{Sec.~\ref}
\newcommand{\Fref}{Fig.~\ref}

\newcommand{\rmd}{{\rm d}}

\def\diff#1#2{{{\rmd #1} \over {\rmd #2}}}

\title{Optimization search effort over the control landscapes for open quantum systems with Kraus-map evolution}
\author{Anand Oza\footnote{Department of Chemistry, Princeton University, Princeton, NJ 08544, USA} , Alexander Pechen$^*$, Jason Dominy\thanks{Program in Applied and Computational Mathematics, Princeton University, Princeton, NJ 08544, USA},\\ Vincent Beltrani$^*$, Katharine Moore$^*$ and Herschel Rabitz$^*$}
\date{May 5, 2009}

\begin{document}
\maketitle

\begin{abstract}
A quantum control landscape is defined as the expectation value of a target observable $\Theta$ as a function of the control variables. In this work control landscapes for open quantum systems governed by Kraus map evolution are analyzed. Kraus maps are used as the controls transforming an initial density matrix $\rho_{\rm i}$ into a final density matrix to maximize the expectation value of the observable $\Theta$. The absence of suboptimal local maxima for the relevant control landscapes is numerically illustrated. The dependence of the optimization search effort is analyzed in terms of the dimension of the system $N$, the initial state $\rho_{\rm i}$, and the target observable $\Theta$. It is found that if the number of nonzero eigenvalues in $\rho_{\rm i}$ remains constant, the search effort does not exhibit any significant dependence on $N$. If $\rho_{\rm i}$ has no zero eigenvalues, then the computational complexity and the required search effort rise with $N$. The dimension of the top manifold (i.e., the set of Kraus operators that maximizes the objective) is found to positively correlate with the optimization search efficiency. Under the assumption of full controllability, incoherent control modelled by Kraus maps is found to be more efficient in reaching the same value of the objective than coherent control modelled by unitary maps. Numerical simulations are also performed for control landscapes with linear constraints on the available Kraus maps, and suboptimal maxima are not revealed for these landscapes.
\end{abstract}

\section{Introduction}
The general goal of quantum control is to apply a suitable external field to a system in order to maximize the expectation value of a target operator. If the system under control is isolated from the environment, then the dynamics are coherent and described by a unitary transformation, as appears in coherent control~\cite{walmsely1,butkov,tannor,judson,warren,rice,rabitz,shapiro,dantus}. In practice, all real systems are open and interact with the environment in some fashion, so that the dynamics of the system will have some incoherent component. Such incoherent control through interaction with the environment~\cite{pechen,pechen2,pechen:ice3,romano,accardi,linington1,fu} or through quantum measurements~\cite{mendes1,mandilara,pechen3,roa,feng,sugny,feng2} can be beneficial in many cases (e.g., in creation of thermal beams of metastable noble gases~\cite{ding}, in quantum computing with mixed states~\cite{tarasov}, or in the modification~\cite{mizel} of Grover's algorithm to extend the capabilities of the original unitary scheme).

In this paper, we consider the most general class of physically-allowed state transformations of controlled open quantum systems. These transformations are represented by Kraus maps~\cite{kraus} providing a kinematic description of incoherent control. Embedded in these maps is information about the system and environment, both of which may be subject to control. A control action determines the system's evolution with a Kraus map $\Phi$, which transforms an initial system state $\rho_{\rm i}$ into the evolved final state $\rho_{\rm f}=\Phi(\rho_{\rm i})$. The final state $\rho_{\rm f}$ determines the expectation value $J[\Phi]:=\langle\Theta\rangle=\Tr[\Phi(\rho_{\rm i})\Theta]=\Tr[\rho_{\rm f}\Theta]$ of a target Hermitian operator $\Theta$ representing a desired physical property to be optimized. The corresponding control goal is formulated as follows: given an initial state $\rho_{\rm i}$ and a target observable $\Theta$, find a Kraus map $\Phi_{\rm opt}$ that transforms $\rho_{\rm i}$ into a state maximizing the expectation value, i.e., such that $J[\Phi_{\rm opt}]=\max\limits_\Phi J[\Phi]$. The set of all Kraus maps for a given quantum system forms a complex Stiefel manifold to formulate the control goal as a nonlinear problem of maximizing the objective function $J$ over the Stiefel manifold.

As shown in~\cite{rongwu1}, for any desired final state $\rho_{\rm f}$ there exists a Kraus map $\Phi_{\rho_{\rm f}}$ that transforms all initial states $\rho_{\rm i}$ into $\rho_{\rm f}$, i.e., such that $\Phi_{\rho_{\rm f}}(\rho_{\rm i}) = \rho_{\rm f}$ for any $\rho_{\rm i}$. If $|\psi\rangle$ is an eigenvector of the target operator $\Theta$ that corresponds to the maximal eigenvalue $\theta_{\rm max}$, then the expectation $\Tr[\rho_{\rm f}\Theta]$ is maximized by the final state $\rho_{\rm f}=\rho_{{\rm f},\psi}=|\psi\rangle\langle\psi|$ and therefore the objective $J[\Phi]=\Tr[\Phi(\rho_{\rm i})\Theta]$ is maximized, e.g., by the optimal map $\Phi=\Phi_{\rho_{{\rm f},\psi}}$. The corresponding maximum objective value is $J_{\rm max}=\theta_{\rm max}$. Thus, the ability to generate dynamically arbitrary Kraus maps for an open quantum system implies its complete state-to-state controllability and, in particular, complete controllability for the objectives of the form $J=\langle\Theta\rangle$. In contrast, a closed quantum system controlled by unitary dynamics has restricted state controllability; if $\rho_{\rm i}$ and $\rho_{\rm f}$ do not have the same eigenvalue spectrum, there does not exist a unitary transformation $U$ such that $U\rho_{\rm i} U^{\dagger} = \rho_{\rm f}$. The maximum attained value for the objective $J=\langle\Theta\rangle$ in this case will generally be less than $\theta_{\rm max}$.

The quantum control landscape is defined by $J=\langle\Theta\rangle$ as a function of the control variables. The ability to successfully use a gradient or other local algorithm for maximization of the objective function depends on the existence or the absence of suboptimal local maxima. If local maxima exist, a local algorithm could get stuck at such points, and for this reason, we refer to suboptimal local maxima as \lq\lq traps;\rq\rq the presence of local saddle points should not serve as traps. In the case of coherent laser control, the landscape is known to be trap free~\cite{lasernotrap,hsieh1}.

A detailed analysis of the control landscapes for incoherent control of open two-level quantum systems was performed~\cite{pechen1}, where the absence of traps for these landscapes was proven. Arbitrary multi-level systems were considered in \cite{wu1}, where it was shown that no suboptimal traps exist for the control landscapes for any finite-level open quantum system. In addition, a high-dimensional submanifold of optimal controls was found. As in the case of coherent control, these results on the absence of traps and the multi-dimensionality of the global optimum manifold provide a theoretical foundation for the empirical fact that it is relatively easy to find optimal solutions even in the presence of an environment.

The absence of traps in control landscapes for both closed and open quantum systems implies that the search using a local algorithm will eventually reach a global optimum solution. However, the absence of traps does not specify the efficiency of optimization procedure and the search effort needed to reach the solution. The efficiency of the optimization procedure to find an optimal control, which is of practical importance due to limitations on computer time in simulations and laboratory resources in experiments, is determined by the local features of the control landscape as well as its topological characteristics. Different trap-free control landscapes can exhibit different degrees of search complexity. The prior relevant theoretical landscape analyses~\cite{pechen1,wu1} for incoherent control of open quantum systems did not describe the dependence of efficiency on the key parameters of the control problem: the dimension of the system $N$ and the eigenvalues of the target operator $\Theta$ and the initial state $\rho$. For closed quantum systems, a theoretical analysis of the computational complexity of coherent control landscapes was performed~\cite{chakra1,chakra2} along with a numerical analysis of the search effort using gradient, genetic and simplex algorithms~\cite{moore08,riviello08}. The results indicate that the search effort scales weakly, or possibly independently, with the dimension of the system $N$.

This paper presents a numerical analysis, with a gradient algorithm, of the search effort for incoherent control of open quantum systems. The analysis lends insight into the topological and structural characteristics of the corresponding quantum control landscapes. It shows that the search effort for driving a pure state into another pure state with Kraus maps remains relatively constant as the dimension $N$ of the system increases, and this behaviour is qualitatively similar to the scaling behavior of the search effort for closed systems~\cite{moore08,riviello08}. A more general result is established for arbitrary, not necessarily pure, initial states: the search effort is essentially determined by the number of nonzero eigenvalues of the initial state $\rho$, and not by the dimension of the system $N$. Thus, when the number of non-zero eigenvalues of the initial state remains constant, the search effort does not depend on $N$. At the extreme of driving a mixed state with no zero eigenvalues into a pure state the search effort increases with the dimension of the system. The detailed analysis shows that the search effort is sensitive to the eigenstructure of the initial state $\rho$ and the target operator $\Theta$; specifically, the degeneracies of the zero eigenvalue of $\rho$ and of the maximal eigenvalue of $\Theta$ positively correlate with the search efficiency, so that higher values of these degeneracies require less optimization search effort and correspond to a more efficient search. Further, comparative analysis of incoherent and coherent control shows that incoherent control under the full controllability assumption is a more efficient process than coherent control, indicating that the additional control freedom afforded by incoherent control can decrease the complexity of the problem. Finally, an analysis of control landscapes with linear constraints on the control variables is performed, and it does not reveal the presence of suboptimal traps even for a large number of independent constraints. Use of Kraus maps for modelling the controlled evolution of the system in this paper greatly simplifies computations as it does not require solving the dynamical evolution equations. Analysis of the scaling properties of the search effort for dynamical optimization of open quantum systems remains as an issue for future study that can be performed using various specific models for the system and the environment~\cite{grace1,alv}.

The paper is organized as follows. \Sref{genform1} describes the general theoretical framework for the kinematic analysis of incoherent control of multilevel open quantum systems. The expressions for the gradient and Hessian of the objective function $J$ are derived in \sref{gradHess}. \Sref{numresult} contains the results of the numerical simulations. Section~4.1 describes the details of the optimization procedure, and section~4.2 discusses the distribution of the objective values for randomly generated controls. \Sref{trapfree1} computationally demonstrates the absence of traps in the control landscape for a five-level quantum system. \Sref{depN} shows the dependence of optimization efficiency on the dimension of the quantum system, and \sref{deprho} examines the dependence of the computational complexity on the degeneracy structure of the eigenvalues of the initial state and target observable. \Sref{unitcomp} compares the computational efficiency of coherent and incoherent control. Optimization over constrained landscapes is investigated in \sref{LinConst}. Concluding remarks are given in \sref{conclusions}.

\section{Formulation of control for arbitrary $N$-level systems}\label{genform1}
In this section the evolution of controlled $N$-level open quantum systems is modelled by Kraus maps. As background, first the common formulation of the objective function in terms of Kraus operators is provided. Then the control problem is  reformulated as optimization over a suitable Stiefel manifold; this representation is used in the subsequent numerical analysis.

\subsection{Kraus maps}
Let ${\cal M}_N$ be the vector space of $N\times N$ complex matrices, with identity matrix ${\mathbb I}_N$. The density matrix $\rho \in {\cal M}_N$ of an $N$-level quantum system is a positive semidefinite (and therefore Hermitian) matrix with unit trace, $\Tr \rho = 1$. A linear map $\Phi: {\cal M}_N \rightarrow {\cal M}_N$ is {\it positive} if $\Phi(M) \geq 0$ for any $M \in {\cal M}_N$ such that $M \geq 0$. The most general evolution transformations of density matrices are given by linear Kraus maps $\Phi: {\cal M}_N \rightarrow {\cal M}_N$, which are defined by the following two properties:
\begin{itemize}
\item Complete positivity: For any integer $n$, the map $\Phi \otimes {\mathbb I}_n$ acting on ${\cal M}_N \otimes {\cal M}_n$ is positive, where $\otimes$ denotes the Kronecker product.
\item Trace preserving: For any $M \in {\cal M}_N$, $\Tr \Phi(M) = \Tr M$.
\end{itemize}

Any Kraus map $\Phi$ can be written in the Kraus operator-sum representation (OSR) form
\begin{equation}\label{krausdec}
\Phi(\rho) = \sum_{i=1}^l K_{i}\rho K_{i}^{\dagger},
\end{equation}
and the trace preservation condition implies for the Kraus operators $K_{i} \in {\cal M}_N$ that the relation is satisfied
\begin{equation}\label{const-1}
\sum_{i=1}^l K_{i}^{\dagger}K_{i} = {\mathbb I}_N.
\end{equation}
There exist many equivalent operator-sum representations of the same Kraus map. In particular, as shown in~\cite{choi1}, for any OSR with $l > N^2$ Kraus operators there exists an equivalent OSR with no more that $N^2$ Kraus operators. Thus, we only need to consider the OSR with $l=N^2$ Kraus operators (some of the Kraus operators can be zero matrices). Even for $l=N^2$ the decomposition~(\ref{krausdec}) is not unique. Indeed, let ${\cal U}(n)$ be the set of $n\times n$ unitary matrices, and let $U \in {\cal U}(N^2)$ be a unitary matrix with matrix elements $u_{ij}$. Define a new set of Kraus operators by the relation
\begin{equation}\label{equivrep1}
\tilde{K}_j = \sum_{i=1}^{N^2} u_{ji}K_i, \qquad 1 \leq j \leq N^2. \nonumber
\end{equation}
Then $\sum_{i=1}^{N^2} \tilde{K}_{i}^{\dagger}\tilde{K}_{i}={\mathbb I}_N$ and $\Phi(\rho) = \sum_{i=1}^{N^2} K_{i}\rho K_{i}^{\dagger} = \sum_{i=1}^{N^2} \tilde{K}_{i}\rho \tilde{K}_{i}^{\dagger}$ for any $\rho$. Therefore the two sets of Kraus operators $\{K_{i}\}$ and $\{\tilde{K}_i \}$ provide two equivalent representations of the same Kraus map.

\subsection{The objective function: formulation in terms of Kraus operators}
The optimization goal in quantum control is to maximize the objective function $J = \langle \Theta\rangle\equiv\Tr[\Phi(\rho)\Theta]=\Tr[\rho_{t_{\rm f}}\Theta]$, where $\Theta$ is some target Hermitian operator, $\langle \cdot \rangle$ denotes the expectation value at the final time $t_{\rm f}$, and $\rho_{t_{\rm f}}$ is the state of the system at the final time, evolved under controls from some initial state $\rho= \rho_{t_0}$. The Kraus operators $\{K_{i}\}\equiv\{K_{i}(t_{\rm f},t_0)\}$ describe the generally non-unitary evolution $\Phi$ of the initial density matrix $\rho$ at time $t_0$ into a density matrix $\rho_{t_{\rm f}}$ at time $t_{\rm f}$, such that $\rho_{t_{\rm f}}=\Phi(\rho)= \sum_{i=1}^{N^2} K_{i}\rho K_{i}^{\dagger}$. They contain the information about the system-environment interaction, all control field interactions, and the state of the environment which also can be used as a control. Hence, $J$ is a function of the Kraus operators
\begin{equation}\label{J-1}
J[K_1,\dots,K_{N^2}] = \Tr \Bigl[\sum_{i=1}^{N^2}K_{i}\rho K_{i}^{\dagger}\Theta\Bigr],
\end{equation}
and the control goal can be formulated as a constrained optimization problem: given $\rho$ and $\Theta$, maximize $J$ over all sets of operators $\{K_{i}\}$ that satisfy the constraint (\ref{const-1}).

For the remainder of the paper, we will take $\rho$ and $\Theta$ to be simultaneously diagonal. Indeed, we can always choose a basis in which $\Theta$ is diagonal, and write $\rho$ and $\{K_{i}\}$ in this basis. Since $\rho$ is Hermitian, there exists a unitary matrix $\Omega$ such that $\rho = \Omega \sigma\Omega^{\dagger}$, where $\sigma$ is a diagonal matrix. Then the objective function~(\ref{J-1}) takes the form
$J = \Tr \left[\sum_{i=1}^{N^2} \tilde{K_{i}}\sigma\tilde{K}_{i}^{\dagger}\Theta\right]$, where $\tilde{K_{i}} = K_{i}\Omega$. The new Kraus operators $\{\tilde{K_{i}}\}$ also satisfy the constraint (\ref{const-1}) and the objective function is equivalently represented as a function of $\tilde K_i$ with simultaneously diagonal matrices $\sigma$ and $\Theta$.

\subsection{The objective function: formulation in terms of Stiefel manifolds}\label{stiefform1}
The above formulation can be expressed more succinctly in terms of the Stiefel manifold \cite{stiefel1}. Let ${\cal M}(n,k,\mathbb{F})$ be the set of $n\times k$ matrices with matrix elements in the field $\mathbb{F}$ of real or complex numbers (i.e., $\mathbb{F} = \mathbb{R}$ or $\mathbb{F} = \mathbb{C}$). The Stiefel manifold is defined as
\[
V_k(\mathbb{F}^n) = \{S \in {\cal M}(n,k,\mathbb{F}): S^{\dagger}S = {\mathbb I}_k\}.
\]
The manifold $V_k(\mathbb{F}^n)$ is called a real (resp., complex) Stiefel manifold if $\mathbb{F} = \mathbb{R}$ (resp., $\mathbb{F} = \mathbb{C}$). Given a Kraus map $\Phi$ and a set of Kraus operators $\{K_{i}\}$, we form the corresponding $N^3\times N$ Stiefel matrix $S$ as follows:
\begin{equation}\label{stiefel-1}
S = \pmatrix{K_{1} \cr K_{2} \cr \vdots \cr K_{N^2}}.
\end{equation}
The constraint $(\ref{const-1})$ can be expressed as the equality $S^{\dagger}S = {\mathbb I}_N$, which defines the complex Stiefel manifold ${\cal S} = V_N(\mathbb{C}^{N^3})$. Furthermore, the objective function $(\ref{J-1})$ can be written as a function of the Stiefel matrix $S$
\begin{equation}\label{J-2}
J(S) = \Tr \left[S\rho S^{\dagger}({\mathbb I}_{N^2}\otimes\Theta)\right],
\end{equation}
The control goal in this formulation is to maximize the objective function (\ref{J-2}) over the Stiefel manifold ${\cal S}$. Note that the objective function~(\ref{J-2}) is by construction real valued for any initial density matrix $\rho$ and for any Hermitian target operator $\Theta$.

We now address the non-uniqueness of the Kraus operator parametrization in terms of the Stiefel manifold. Let ${\cal W} = \{U \otimes {\mathbb I}_N: U \in {\cal U}(N^2)\}$. It is straightforward to verify that $\forall S \in {\cal S}$ and $\forall W \in {\cal W}$ holds $\tilde{S} \equiv WS \in {\cal S}$. If $\{K_i\}$ and $\{\tilde{K}_i\}$ are two sets of Kraus operators that determine two Stiefel matrices $S$ and $\tilde{S}$ through~(\ref{stiefel-1}), then they define the same Kraus map and are related by the equality~(\ref{equivrep1}) if and only if $\exists W \in {\cal W}$ such that $\tilde S = WS$. Thus, equivalent parametrizations of the same Kraus map correspond to Stiefel matrices related by $\tilde S = WS$ with some $W \in {\cal W}$. This property implies the invariance of the objective function under ${\cal W}$-transformations, $J(S) = J(WS)$ for any $W \in {\cal W}$, and will be used in \sref{LinConst} for analyzis of the search effort for optimization of $J$ with additional constraints on the available Kraus operators.

The Stiefel manifold $V_k(\mathbb F^n)$ can also be defined as the set of orthonormal $k$-frames in $\mathbb{F}^n$~\cite{Hatcher2002}. In this way, the Stiefel manifold $\cal S$ can be specified as the set of ordered $N$-tuples $X_1,\dots,X_N \in \mathbb{C}^{N^3}$ such that $\langle X_i,X_j\rangle = \delta_{ij}$, where $\delta_{ij}$ is the Kronecker delta symbol and $\langle \cdot,\cdot\rangle$ denotes the inner product in $\mathbb{C}^{N^3}$. In the remainder of the manuscript, the notation $\langle \cdot,\cdot\rangle$ will be used for inner products in several appropriate different spaces (namely, standard inner products in $\mathbb{C}^{N^3}$ and in $\mathbb{C}^{N^2}$, and real Hilbert-Schmidt inner product in $S$ and in the tangent space $T_S{\cal S}$ at $S$). Vector $X_i$ in this representation contains elements of the $i$th column of the Stiefel matrix~(\ref{stiefel-1}) in certain order and can be decomposed in the direct sum
\[
X_i = Y_1^{i} \oplus Y_2^{i} \oplus \cdots \oplus Y_N^{i}
\]
Here each $Y_j^{i} \in \mathbb{C}^{N^2}$, where $1 \leq i,j \leq N$, is a complex vector of length $N^2$ of the form
$Y_j^{i}=\{ (K_1)_{ji},(K_2)_{ji},\dots, (K_{N^2})_{ji}\}$, i.e., components of the vector $Y_j^{i}$ are the $ji$-th matrix elements of all the $N^2$ Kraus operators $K_l$. The orthogonality condition $\langle X_i,X_j\rangle = \delta_{ij}$ implies the relation
\begin{equation}\label{y-relations}
\sum_{k=1}^N \langle Y_k^{i},Y_k^{j}\rangle = \delta_{ij},
\end{equation}
Here $\langle \cdot,\cdot\rangle$ denotes the inner product in $\mathbb{C}^{N^2}$ and should be distinguished from the same notations used above to denote the inner product in $\mathbb{C}^{N^3}$.

The objective function for diagonal matrices $\rho = \sum_{i=1}^N\rho_{i}|i\rangle\langle i|$ and $\Theta = \sum_{j=1}^N\theta_{j} |j\rangle\langle j|$ can be written as
\begin{equation}\label{Yform}
J[\{Y_j^{i}\}] = \sum_{i,j=1}^N \|Y_{j}^{i}\|^2\rho_{i}\theta_{j}.
\end{equation}
It is clear that $\theta_{\rm min} \leq J(S) \leq \theta_{\rm max}$, where $\theta_{\rm min}$ and $\theta_{\rm max}$ are the minimum and maximum eigenvalues of $\Theta$, respectively. Indeed, we have
\[
\theta_{\rm min} \sum_{i,j=1}^N \|Y_{i}^{j}\|^2\rho_{j} \leq J \leq \theta_{\rm max} \sum_{i,j=1}^N \|Y_{i}^{j}\|^2\rho_{j}.
\]
Now, by first summing over $i$ and using~(\ref{y-relations}), and then summing over $j$ and using $\Tr\rho=1$, we have the desired inequalities.

Since the maximal value of the objective function $J$ equals to $\theta_{\rm max}$, the set of optimal controls (i.e., the set of all Stiefel matrices which maximize the objective function) is the manifold ${\cal M}_{\rm max} = \{S \in {\cal S}: J(S) = \theta_{\rm max}\}$. For the case $\Theta = |N\rangle\langle N|$ of special interest, it follows from~(\ref{Yform}) that
\[
{\cal M}_{\rm max} = \left\{\{Y_{j}^{i}\}_{i,j=1}^N: \|Y_{j}^{i}\|^2 =\delta_{jN} \textrm{ for any } i \textrm{ such that } \rho_i\ne 0\right\}.
\]

\section{Gradient and Hessian of $J$}\label{gradHess}
The numerical analysis in section \ref{numresult} uses a gradient algorithm for optimization of the objective function $J(S)$. This algorithm requires solving the equation
\begin{equation}\label{graddiffeq}
\frac{\rmd S}{\rmd\sigma} = {\rm grad }\, J(S).
\end{equation}
Here ${\rm grad}\,J$ is the gradient of the objective function, which induces the corresponding gradient flow on the Stiefel manifold ${\cal S}$ via Eq.~(\ref{graddiffeq}).

\subsection{Gradient of $J$}\label{alg}
We now derive an explicit expression for the gradient. Denote the differential of $J$ at $S \in {\cal S}$ by ${\rmd}_S J: T_S{\cal S} \rightarrow \mathbb{R}$, where $T_S{\cal S}$ is the tangent space at $S$. By the product rule for derivatives
\begin{equation}\label{diffJ2,eq1}
{\rmd}_S J(\delta S) = \Re \Tr\left[(\delta S) \rho S^{\dagger} ({\mathbb I}_{N^2}\otimes\Theta)+S\rho(\delta S)^{\dagger}({\mathbb I}_{N^2}\otimes\Theta)\right]
\end{equation}
where real part $\Re$ is taken since the objective~(\ref{J-2}) is a real function. Since $\Re\Tr A=\Re\Tr A^\dagger$ for any matrix $A$, the second term in the right hand side of~(\ref{diffJ2,eq1}) can be rewritten as $\Re \Tr [({\mathbb I}_{N^2}\otimes\Theta) (\delta S) \rho S^{\dagger}]$ and we get
\begin{eqnarray}\label{diffJ2}
{\rmd}_S J(\delta S) &=& \Re \Tr \left[(\delta S) \rho S^{\dagger} ({\mathbb I}_{N^2}\otimes\Theta)+({\mathbb I}_{N^2}\otimes\Theta) (\delta S) \rho S^{\dagger}\right]  \nonumber\\
&=& \Re \Tr \left[ \rho S^{\dagger} ({\mathbb I}_{N^2}\otimes\Theta)(\delta S) + \rho S^{\dagger}({\mathbb I}_{N^2}\otimes\Theta) (\delta S)\right]  \nonumber \\
&=& 2\Re \Tr \left[ \rho S^{\dagger} ({\mathbb I}_{N^2}\otimes\Theta)(\delta S)\right] \nonumber \\
&=&\langle 2({\mathbb I}_{N^2}\otimes\Theta) S \rho, \delta S\rangle,
\end{eqnarray}
where $\delta S \in T_S{\cal S}$, and $\langle A,B\rangle := \Re\Tr [A^{\dagger}B]$ is the inner product on ${\cal S}$ and $T_S{\cal S}$. By the Riesz Representation Theorem, there exists $X \in T_S{\cal S}$ such that ${\rmd}_SJ(\delta S) = \langle X,\delta S\rangle$ for all $\delta S \in T_S{\cal S}$. The vector $X$ is the {\it gradient} of $J$ at $S$, denoted by ${\rm grad }\, J(S)$.

Since ${\rm grad }\, J(S)$ must lie in $T_S{\cal S}$, it is necessary to remove the component orthogonal to $T_S{\cal S}$ from the vector $2({\mathbb I}_{N^2} \otimes\Theta)S\rho$ appearing in the last line of Eq.~(\ref{diffJ2}). Differentiation of the identity $S^{\dagger}S = {\mathbb I}_N$ gives $S^{\dagger}(\delta S) = -(\delta S)^{\dagger}S$, so $S^{\dagger}(\delta S)$ is skew-Hermitian. This can be rewritten as $\delta S = SB+({\mathbb I}_{N^3}-SS^{\dagger})D$, where $B\in {\cal M}(N,N,\mathbb{C})$ is a skew-Hermitian matrix and $D\in {\cal M}(N^3,N,\mathbb{C})$ is an arbitrary matrix. (Note that $S^{\dagger}(\delta S) = B$, since $S^{\dagger}S = {\mathbb I}_N$). Any $A \in {\cal M}(N^3,N,\mathbb{C})$ can be decomposed as follows:
\[
A =S\frac{1}{2}(S^{\dagger}A+A^{\dagger}S)+S\frac{1}{2}(S^{\dagger}A-A^{\dagger}S)+({\mathbb I}_{N^3}-SS^{\dagger})A.
\]
Let $C = (S^{\dagger}A+A^{\dagger}S)/2$ and $B = (S^{\dagger}A-A^{\dagger}S)/2$, so that
$A = SC+SB+({\mathbb I}_{N^3}-SS^{\dagger})A$. Clearly $C$ is Hermitian and $B$ is skew-Hermitian, so $\langle SC, SB-({\mathbb I}_{N^3}-SS^{\dagger})A\rangle = 0$. Therefore, $SC$ is orthogonal to $T_S{\cal S}$, and hence $(A-SC) \in T_S{\cal S}$. As a result, ${\cal P}_S(A) = A - S(S^{\dagger}A+A^{\dagger}S)/2$ is an orthogonal projector from ${\cal M}(N^3,N,\mathbb{C})$ onto $T_S{\cal S}$, and
\begin{eqnarray}
{\rm grad}\, J(S) &=& 2({\mathbb I}_{N^2}\otimes\Theta) S \rho - S\left[S^{\dagger}({\mathbb I}_{N^2}\otimes\Theta) S \rho+(({\mathbb I}_{N^2}\otimes\Theta) S \rho)^{\dagger}S\right] \nonumber\\
&=&(2{\mathbb I}_{N^3}-SS^{\dagger})({\mathbb I}_{N^2}\otimes\Theta) S \rho - S\rho S^{\dagger}({\mathbb I}_{N^2}\otimes\Theta) S. \nonumber
\end{eqnarray}

\subsection{Hessian of $J$}\label{hessian}
In the analysis thus far, we have only considered ${\rm grad }\, J(S)$, which gives first-order information about $J$. The Hessian gives useful second-order information about the minima, maxima, and saddles of $J$ (where ${\rm grad }\, J(S) = 0$). At such points, the eigenvectors of the Hessian with positive (resp. negative) eigenvalues correspond to directions in which $J$ increases (resp. decreases).

The Hessian of $J$ at $S \in {\cal S}$ acting on $\delta S \in T_S{\cal S}$ is defined as the covariant derivative of ${\rm grad }\, J(S)$ in the direction $\delta S$ \cite{doCarmo1}:
\[
{\rm Hess }\, J(S): T_S {\cal S} \rightarrow T_S {\cal S}, \qquad {\rm Hess }\, J(S)(\delta S) = \nabla_{\delta S}\, {\rm grad }\, J(S).
\]
Covariant differentiation of a function on a vector space is equivalent to taking the ordinary differential. However, ${\cal S}$ is not a vector space. In the following, the strategy will be to take the covariant derivative of ${\rm grad }\, J(S)$ as a function on ${\cal M}(N^3,N,\mathbb{C})$, which is a vector space, and then project this onto ${\cal S}$. Since ${\cal S}$ inherits its inner product from ${\cal M}(N^3,N,\mathbb{C})$, this strategy gives the covariant derivative of ${\rm grad}\, J(S)$ on ${\cal S}$.

We now calculate an expression for the eigenvalues and eigenvectors of the Hessian of $J$ on the critical manifolds. By differentiating ${\rm grad }\, J(S)$ in the direction of $\delta S$, we obtain
\begin{eqnarray}
\overline{\nabla}_{\delta S}\,{\rm grad }\, J(S)&=&{\rmd}_S{\rm grad }\, J(\delta S)=2({\mathbb I}_{N^2}\otimes\Theta) (\delta S)\rho \nonumber \\
&&  -  (\delta S)S^{\dagger}({\mathbb I}_{N^2}\otimes\Theta) S \rho  - S(\delta S)^{\dagger}({\mathbb I}_{N^2}\otimes\Theta) S \rho \nonumber\\
&&-  SS^{\dagger}({\mathbb I}_{N^2}\otimes\Theta)(\delta S)\rho  -  (\delta S)\rho S^{\dagger}({\mathbb I}_{N^2}\otimes\Theta) S \nonumber\\
&&- S\rho(\delta S)^{\dagger}({\mathbb I}_{N^2}\otimes\Theta) S - S\rho S^{\dagger}({\mathbb I}_{N^2}\otimes\Theta)(\delta S),\nonumber
\end{eqnarray}
where $\overline{\nabla}$ denotes the Riemannian connection on ${\cal M}(N^3,N,\mathbb{C})$.
We now project this onto $T_S{\cal S}$. Letting $A = \overline{\nabla}_{\delta S}\, {\rm grad }\, J(S)$ gives
\[
{\rm Hess }\, J(S)(\delta S) = \nabla_{\delta S}\,{\rm grad }\, J(S) = P_{\cal S}(\overline{\nabla}_{\delta S}\,{\rm grad }\, J(S)) = A - \frac{1}{2}S(S^{\dagger}A+A^{\dagger}S).
\]
With some algebra, this expression can be reduced to
\begin{eqnarray}
 {\rm Hess }\,J(S)(\delta S) &=&2({\mathbb I}_{N^2}\otimes\Theta) (\delta S)\rho - (\delta S)S^{\dagger} ({\mathbb I}_{N^2}\otimes\Theta) S\rho - (\delta S)\rho S^{\dagger}({\mathbb I}_{N^2}\otimes\Theta) S \nonumber \\
&&+ \frac{1}{2}[SS^{\dagger}(\delta S)S^{\dagger}({\mathbb I}_{N^2}\otimes\Theta) S\rho + SS^{\dagger}(\delta S)\rho S^{\dagger}({\mathbb I}_{N^2}\otimes\Theta) S \nonumber \\
&&-2SS^{\dagger}({\mathbb I}_{N^2}\otimes\Theta) (\delta S)\rho - 2S\rho(\delta S)^{\dagger}({\mathbb I}_{N^2}\otimes\Theta) S \nonumber \\
&&+S\rho S^{\dagger}({\mathbb I}_{N^2}\otimes\Theta) S(\delta S)^{\dagger}S + SS^{\dagger}({\mathbb I}_{N^2}\otimes\Theta) S\rho(\delta S)^{\dagger}S]. \nonumber
\end{eqnarray}
Combining the first two terms in the square brackets gives
\begin{eqnarray}
&&SS^{\dagger}(\delta S)[S^{\dagger}({\mathbb I}_{N^2}\otimes\Theta) S\rho+\rho S^{\dagger}({\mathbb I}_{N^2}\otimes\Theta) S]\nonumber\\ &&\qquad\qquad\qquad= SS^{\dagger}(\delta S)
[2S^{\dagger}({\mathbb I}_{N^2}\otimes\Theta) S\rho - S^{\dagger}{\rm grad }\, J(S)] \nonumber \\
&&\qquad\qquad\qquad=2SS^{\dagger}(\delta S)S^{\dagger}({\mathbb I}_{N^2}\otimes\Theta) S\rho, \nonumber
\end{eqnarray}
since ${\rm grad }J(S) = 0$ at a critical point. Combining the last two terms in the square brackets gives
\begin{eqnarray}
 &&[S\rho S^{\dagger}({\mathbb I}_{N^2}\otimes\Theta) S+ SS^{\dagger}({\mathbb I}_{N^2}\otimes\Theta) S\rho]((\delta S)^{\dagger}S) \nonumber \\ &&\qquad\qquad\qquad= [2({\mathbb I}_{N^2}\otimes\Theta) S\rho - {\rm grad }\, J(S)]((\delta S)^{\dagger}S)\nonumber \\
&&\qquad\qquad\qquad= 2({\mathbb I}_{N^2}\otimes\Theta) S\rho(\delta S)^{\dagger}S. \nonumber
\end{eqnarray}
As a result, we have
\begin{eqnarray}
 {\rm Hess }\, J(S)(\delta S) &=& 2({\mathbb I}_{N^2}\otimes\Theta) (\delta S)\rho - (\delta S)S^{\dagger} ({\mathbb I}_{N^2}\otimes\Theta) S\rho - (\delta S)\rho S^{\dagger}({\mathbb I}_{N^2}\otimes\Theta) S \nonumber \\
&&- SS^{\dagger}({\mathbb I}_{N^2}\otimes\Theta) (\delta S)\rho + SS^{\dagger}(\delta S)S^{\dagger}({\mathbb I}_{N^2}\otimes\Theta) S\rho \nonumber \\
&& - S\rho(\delta S)^{\dagger}({\mathbb I}_{N^2}\otimes\Theta) S
+ ({\mathbb I}_{N^2}\otimes\Theta) S\rho(\delta S)^{\dagger}S. \nonumber
\end{eqnarray}

\section{Numerical assessment of optimization efficiency for landscapes without constraints}\label{numresult}
This section presents numerical simulations, including (a) an empirical demonstration of the absence of suboptimal traps in the control landscape, (b) an analysis of the dependence of optimization efficiency on the dimension of the system $N$, target operator $\Theta$, and initial state $\rho$, and (c) a comparison between coherent and incoherent control.

\subsection{The optimization procedure}
We now describe the procedure for the numerical analysis of the controlled excursions over the landscapes without constraints on the controls. First, an adapted version of the algorithm in~\cite{mezzadri1} is used to randomly generate an initial Stiefel matrix $S_0$ with a uniform distribution on the Stiefel manifold ${\cal S}$. After the initial Stiefel matrix is generated, the Runge-Kutta method built into MATLAB is used to solve Eq.~(\ref{graddiffeq}) with the initial condition $S(0)=S_0$. The method relies on using a variable step size. The tolerances in the differential equation solver are set so that $\|S^{\dagger}S-{\mathbb I}_N\| < 2\times 10^{-4}$ at any given point in the trajectory. Integration is terminated when $J(S)>(\theta_{\rm max}-0.01)$.

The efficiency of the optimization procedure is measured by the two parameters: (1) the number $\tau$ of $\sigma$-steps taken by the differential equation solver in MATLAB to reach the objective value $J>(\theta_{\rm max}-0.01)$ and (2) the path length $\lambda$ taken to get there. A higher number $\tau$ of $\sigma$-steps corresponds to a more difficult optimization problem. Given the number $\tau$ of $\sigma$-steps, the path length $\lambda$ is defined as
\begin{equation}
 \lambda= \sum_{i=0}^{\tau-1} ||S(i+1)-S(i)||,
\end{equation}
where $\|S\|=\sqrt{\langle S,S\rangle}$ is the norm on ${\cal S}$. Similarly, a large value of $\lambda$ corresponds to a convoluted trajectory through ${\cal S}$ and indicates an inefficient optimization.

To ensure statistical uniformity, for some simulations an average was performed over the initial state with a uniform distribution. Uniform sampling on the space of diagonal density matrices is implemented as follows. Let ${\cal V}_n^+$ be the standard simplex, i.e., the set of all vectors $z = (z_1,\dots,z_n) \in \mathbb{R}^n$ such that $z_i \geq 0$ and $\sum_{i=1}^n z_i = 1$. Let $x_i = -\log(a_i)$ where $a_i$ is uniformly distributed on $[0,1]$, so $x_i$ are exponentially distributed with parameter 1. Now let
\[
y_i = \frac{x_i}{x_1+\dots+x_n}, \qquad 1 \leq i \leq n.
\]
Then the random vector $y = (y_1,\dots,y_n)$ is uniformly distributed on the simplex ${\cal V}_n^+$~\cite{devroy1} and the diagonal density matrix $\rho$ with matrix elements $\rho_{ii} = y_i$ is uniformly distributed.

\subsection{The statistical distribution of the objective for randomly generated controls}
\begin{figure}[h]\center
\includegraphics[scale=0.35]{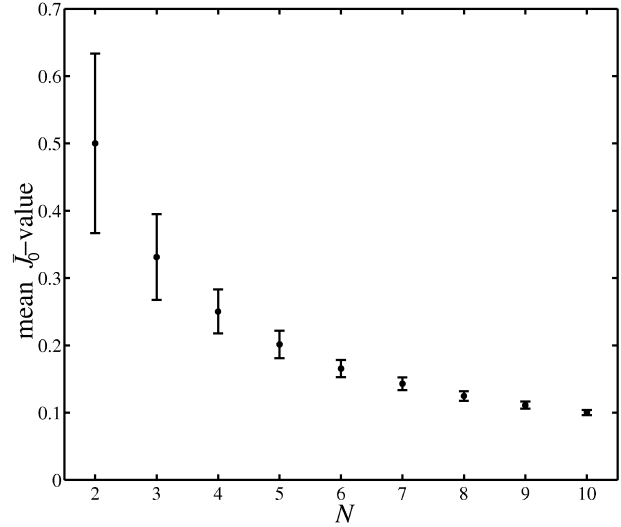}
\caption{The mean value of the objective function ${J_0}(S_0,\rho) = \Tr[S_0\rho S_0^{\dagger}({\mathbb I}_{N^2} \otimes\Theta)]$, where $S_0$ has a uniform distribution on the Stiefel manifold, $\rho$ has a uniform distribution over the set of diagonal density matrices, and $\Theta = |N\rangle\langle N|$.  500 samples were taken for every point $N$. The error bars show the standard deviation for each $N$.}
\label{Jval}
\end{figure}

In practical optimization of the objective function, either in the laboratory or through simulations with a numerical algorithm, the initial control is usually randomly generated. As the Stiefel matrices serve as the controls, we first analyze the distribution of the objective value for randomly generated initial Stiefel matrices. \Fref{Jval} shows the mean value $\bar J_0$ of the objective function $J_0(S_0,\rho) = \Tr[S_0\rho S_0^{\dagger}({\mathbb I}_{N^2}\otimes\Theta)]$ for $\Theta=|N\rangle\langle N|$ as a function of the system dimension $N$ for a uniform distribution of the initial Stiefel matrix $S_0$ and uniform distribution of the initial diagonal density matrices $\rho$. For this case the mean value $\bar J_0$ equals to $1/N$. To understand this result, let $|1\rangle,\dots,|N\rangle$ be an orthonormal basis in the Hilbert space of the system such that $|N\rangle\equiv|\Psi\rangle$ is the target state. The uniform generation of the Stiefel matrix $S_0$ and initial density matrix $\rho$ does not have a preferred state and thus preserves the symmetry between the states $|1\rangle,\dots,|N\rangle$. Therefore, in the final density matrix $\rho'$ obtained by applying to $\rho$ the Kraus map associated to Stiefel matrix $S_0$, the averaged (over uniform distributions of $\rho$ and $S_0$) population $p_i$ of each of these states will be the same for all $i$. Since $\sum\limits_{i=1}^N p_i=1$ and $p_1=p_2=\dots=p_N$, we have $p_i = 1/N$ for each $i$. Hence, the mean value of $J_0 = \Tr[S_0\rho S_0^{\dagger}({\mathbb I}_{N^2}\otimes\Theta)]\equiv p_N$ will be $1/N$ for $\Theta = |\Psi\rangle\langle\Psi|$ being a projector onto the target state $|\Psi\rangle$.

\begin{figure}\center
\includegraphics[scale=0.35]{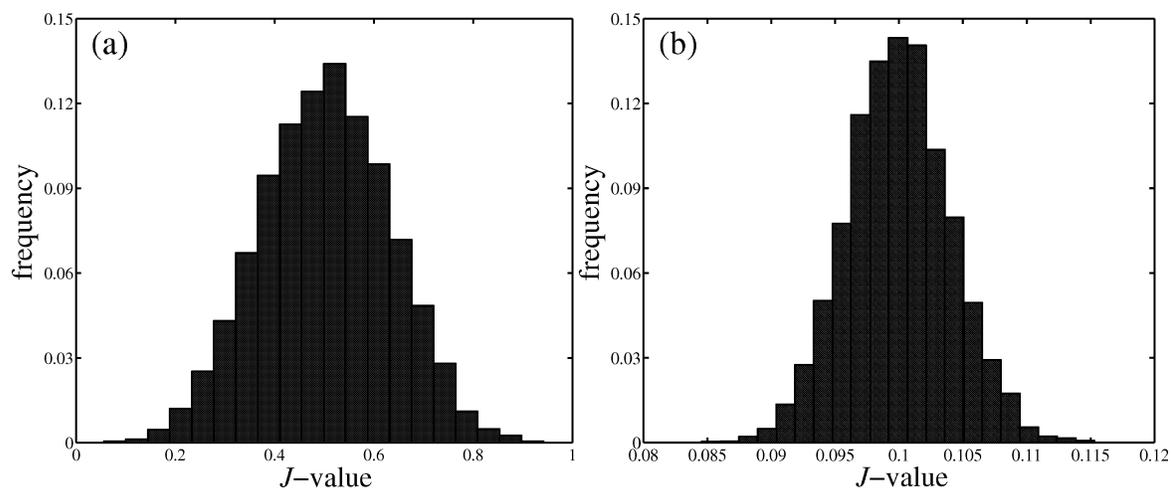}
\caption{The distribution of the values of the objective function $J_0(S_0,\rho)=\Tr[S_0\rho S_0^{\dagger}({\mathbb I}_{N^2} \otimes\Theta)]$ for $N=2$ [subplot (a)] and $N=10$ [subplot (b)]. The initial Stiefel matrix $S_0$ is uniformly distributed on the Stiefel manifold, $\rho$ is uniformly distributed on the set of diagonal density matrices, and the target operator has the form $\Theta = |N\rangle\langle N|$. $10^4$ samples were taken to produce the statistics for each plot.}
\label{Jdist}
\end{figure}

\Fref{Jval} shows the decrease in the expected initial value of the objective function along with a decrease in the standard deviation with increasing system dimension $N$. \Fref{Jdist} presents the detailed form of the distributions for the cases $N=2$ and $N=10$, respectively shown in 2a and 2b, with a uniform distribution of $S_0$ on the Stiefel manifold and a uniform distribution of $\rho$ on the set of diagonal matrices. In this figure, the distributions of the values of the objective function $J_0$ are produced using $10^4$ randomly selected pairs of $S_0$ and $\rho$. The results agree with the natural expectation that the efficiency of a randomly choosen control decreases with increasing complexity of the system. The figure also shows that as $N$ rises the distribution of the objective values becomes more concentrated around the mean value. An open issue is to obtain an analytical expression for the distribution of the initial objective value $J_0(S_0,\rho)$.

\subsection{Absence of suboptimal traps}\label{trapfree1}
Let $X$ be a topological space and $f: X \rightarrow \mathbb{R}$. The function $f$ is said to have a local maximum at $x_0 \in X$ if there exists an open neighborhood of $x_0$, $U_{x_0} \subset X$, such that $\forall x \in U_{x_0}$, $f(x) \leq f(x_0)$ and yet there exists some $x_1 \in X$ such that $f(x_1) > f(x_0)$. If $X$ represents the space of all controls and $f: X \rightarrow \mathbb{R}$ is the objective function to be maximized on $X$, then a local maximum of $f$ is called a suboptimal (or false) trap in the control landscape produced by $f$.

\begin{figure}\center
\includegraphics[scale=0.82]{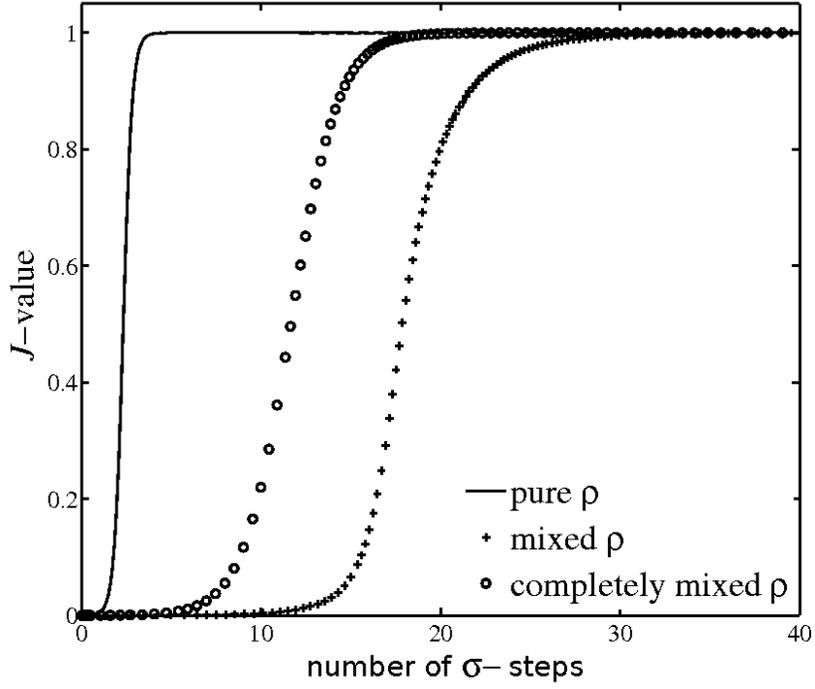}
\caption{The figure shows the value of the objective function $J$ at each step $\sigma$ in the trajectory for a pure initial state $\rho={\rm diag}(1,0,0,0,0)$, a mixed initial state, and a completely mixed initial state $\rho=1/5\cdot{\rm diag}(1,1,1,1,1)$. All three cases correspond to $N=5$ and $\Theta = |5\rangle\langle 5|={\rm diag}(0,0,0,0,1)$. Each trajectory reaches perfect control $J=1$ at the top of the landscape.}
\label{notrapsfig}
\end{figure}

The control landscape for the objective function $J: {\cal S} \rightarrow \mathbb{R}$ defined by~(\ref{J-2}) is known to have no traps~\cite{wu1}. Figure~\ref{notrapsfig}  numerically demonstrates this general fact for a particular five-level quantum system. In the figure, three different initial density matrices are considered: a pure state $\rho = |1\rangle\langle 1|$, a randomly generated mixed state, and a completely mixed state $\rho = {\mathbb I}_5/5$. The control goal is to transform each of these states into the final state $\rho_{\rm f} = |5\rangle\langle 5|$, which maximizes the expectation of the target operator $\Theta = |5\rangle\langle 5|$. As shown in the figure, in each case the gradient algorithm is able to find the control corresponding to the maximum value $J = 1$ of the objective function. The algorithm was not impeded by suboptimal local maxima, for their presence would have caused the algorithm to terminate at $J < 1$. Many other cases showed the same trap free behavior (not shown here).

\subsection{Dependence of the search effort on the dimension of the system}\label{depN}
We now analyze how the dimension $N$ of the controlled system affects the optimization search effort. The goal is to numerically analyze the statistical dependence upon $N$ of the number of steps $\tau$ to reach convergence and the path length $\lambda$. In this section, the target operator $\Theta = |N\rangle\langle N|$ is the projector onto the state $|N\rangle$. To obtain reasonable statistics, for each $N$ we average over 50 simulations of the optimization procedure with randomly (uniformly across ${\cal S}$) generated $S(0)$ and randomly generated initial density matrices $\rho$. We also analyze how the number of zero eigenvalues of $\rho$ (henceforth denoted as $d_0$) affects the scaling of optimization efficiency with $N$.

\begin{figure}\center
\includegraphics[scale=0.85]{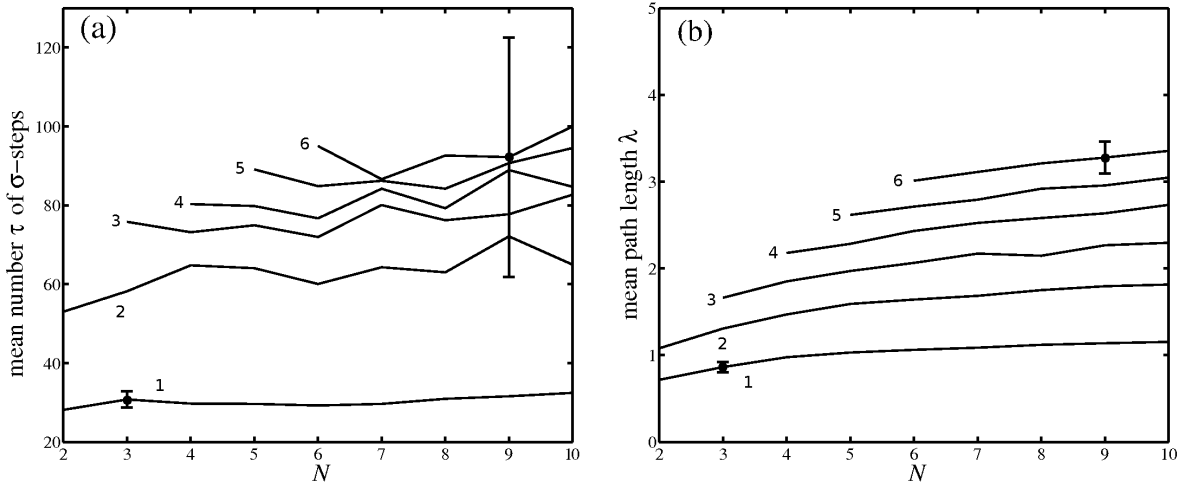}
\caption{Dependence of search efficiency on the dimension of the system $N$ for fixed numbers $N-d_0$ of nonzero eigenvalues of $\rho$. Fifty simulations were performed for each  point, and the average values of the number $\tau$ of $\sigma$-steps and the path length $\lambda$ are plotted in (a) and (b), respectively. On each subplot, the six lines, from bottom to top, correspond to the number of nonzero eigenvalues of the initial density matrix $N-d_0 = $ 1, 2, 3, 4, 5, 6, respectively. The error bars indicate the typical standard deviation of the data for the cases $N-d_0 = 1$ and $N-d_0 = 6$.}
\label{depN0}
\end{figure}

\begin{figure}\center
\includegraphics[scale=0.8]{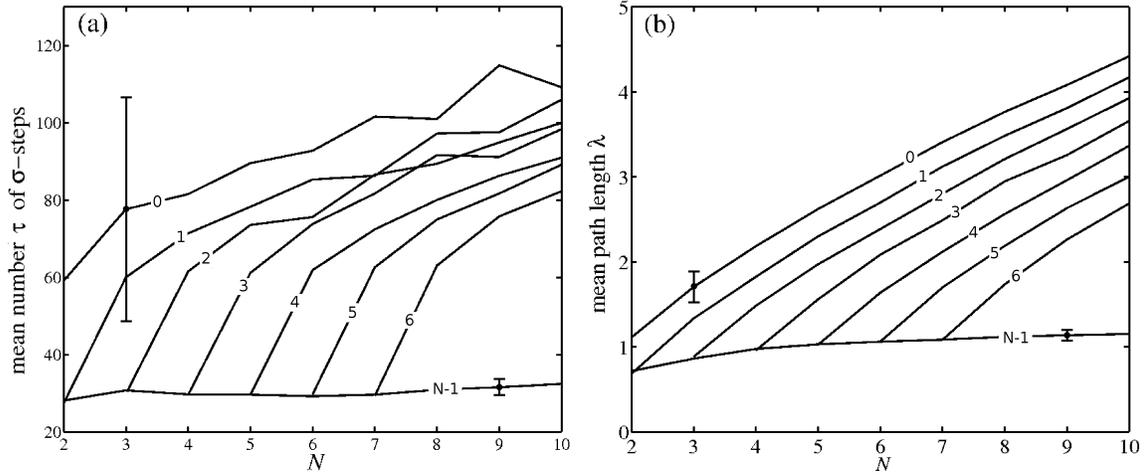}
\caption{Dependence of search efficiency on $N$ for different numbers $d_0$ of zero eigenvalues of $\rho$. Fifty simulations were performed for each  point, and the average values of the number $\tau$ of $\sigma$-steps and the path length $\lambda$ are plotted in (a) and (b), respectively. On each subplot, the eight lines, from top to bottom, correspond to the number of zero eigenvalues of the initial density matrix $d_0 = $ 0, 1, 2, 3, 4, 5, 6, $N-1$, respectively.  The error bars indicate the typical standard deviation of the data for the cases $d_0 = 0$ and $d_0 = N-1$.}
\label{depN1}
\end{figure}

In the case of mixed $\rho$, we change $\rho$ at the start of each individual simulation. Figures~\ref{depN0} and \ref{depN1} plot $\tau$ and $\lambda$ in two different ways in order to illustrate the issues driving the scaling efficiency. For each of the six curves in \Fref{depN0}, the number of nonzero eigenvalues $N-d_0$ of the initial state $\rho$ remains fixed. Each curve labelled by $N-d_0$ corresponds, for example, to the control of a sequence of quantum systems prepared initially in a state at a relatively low temperature, with no population in $d_0$ high eigenstates of the density matrix. Both $\tau$ and $\lambda$ do not show any significant dependence upon $N$. It is clear that for fixed $N$, the search efficiency is greater for larger values of $d_0$. However, increasing $d_0$ for fixed $N-d_0$ does not change the slope of the curves in \Fref{depN0}, showing that the complexity of the search remains relatively insensitive to $N$. The most efficient control problem considered in the figure is the transformation of a pure initial state $\rho = |j\rangle\langle j|$ ($j \neq N$) into a pure final state $\rho_{\rm f} = |N\rangle\langle N|$.

In contrast, for the simulations in \Fref{depN1}, $d_0$ is held fixed for all $N$.  This corresponds, for example, to the control of a sequence of quantum systems with the initial state at ever higher temperature as $N$ rises, producing a large number of populated energy states. Both $\tau$ and $\lambda$ increase quite sharply as $N$ increases, as shown in \Fref{depN1}. It is clear that for fixed $N$, the efficiency of optimization increases as $d_0$ increases. However, as with \Fref{depN0}, increasing $d_0$ does not change the slope of the curves in \Fref{depN1}, showing that the efficiency remains sensitive to $N$. The most inefficient search corresponds to the control goal of transforming a maximum entropy initial state with $d_0 = 0$ to a pure final state with $d_0 = N-1$, which agrees with simple intuition.

The conclusion from Figures~\ref{depN0} and~\ref{depN1} is that when $\Theta$ is a projector, the search efficiency decreases with increasing numbers of {\it nonzero} eigenvalues of $\rho$. The overall dimension of the quantum system $N$ has little effect upon the search efficiency, provided that the number of nonzero eigenvalues of $\rho$ remains fixed. The large standard deviations in both figures are most likely caused by fluctuations of the initial Stiefel matrix $S$ and of the parameters of the initial density matrix $\rho$ not included in the number of zero eigenvalues $d_0$.

The results in \Fref{depN0} have practical relevance. In the laboratory a sequence of quantum systems with increasing $N$ and a roughly fixed small number of populated energy levels can be arranged. Under these conditions, the results shown in \Fref{depN0} indicate that search effort in the laboratory should not be very sensitive to the dimension of the quantum system under control. This behavior is generally consistent with the broad fingings that system and environmental complexity appear to have little effect on the number of iterations to reach successful control in the laboratory.

\subsection{Dependence of the search effort on the degeneracy structure of $\rho$ and $\Theta$}\label{deprho}
\begin{figure}\center
\includegraphics[scale=0.32]{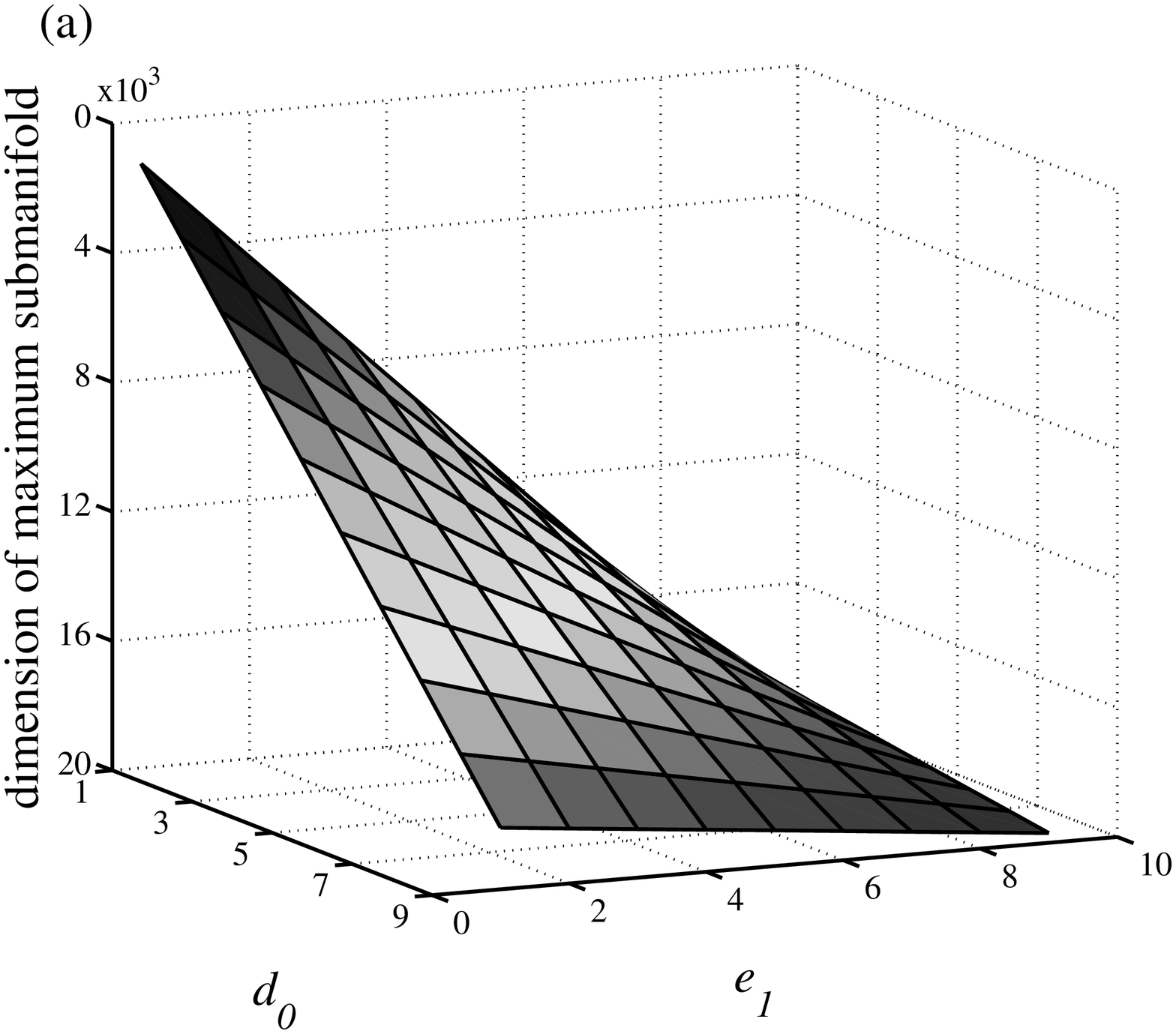}\\
\includegraphics[scale=0.32]{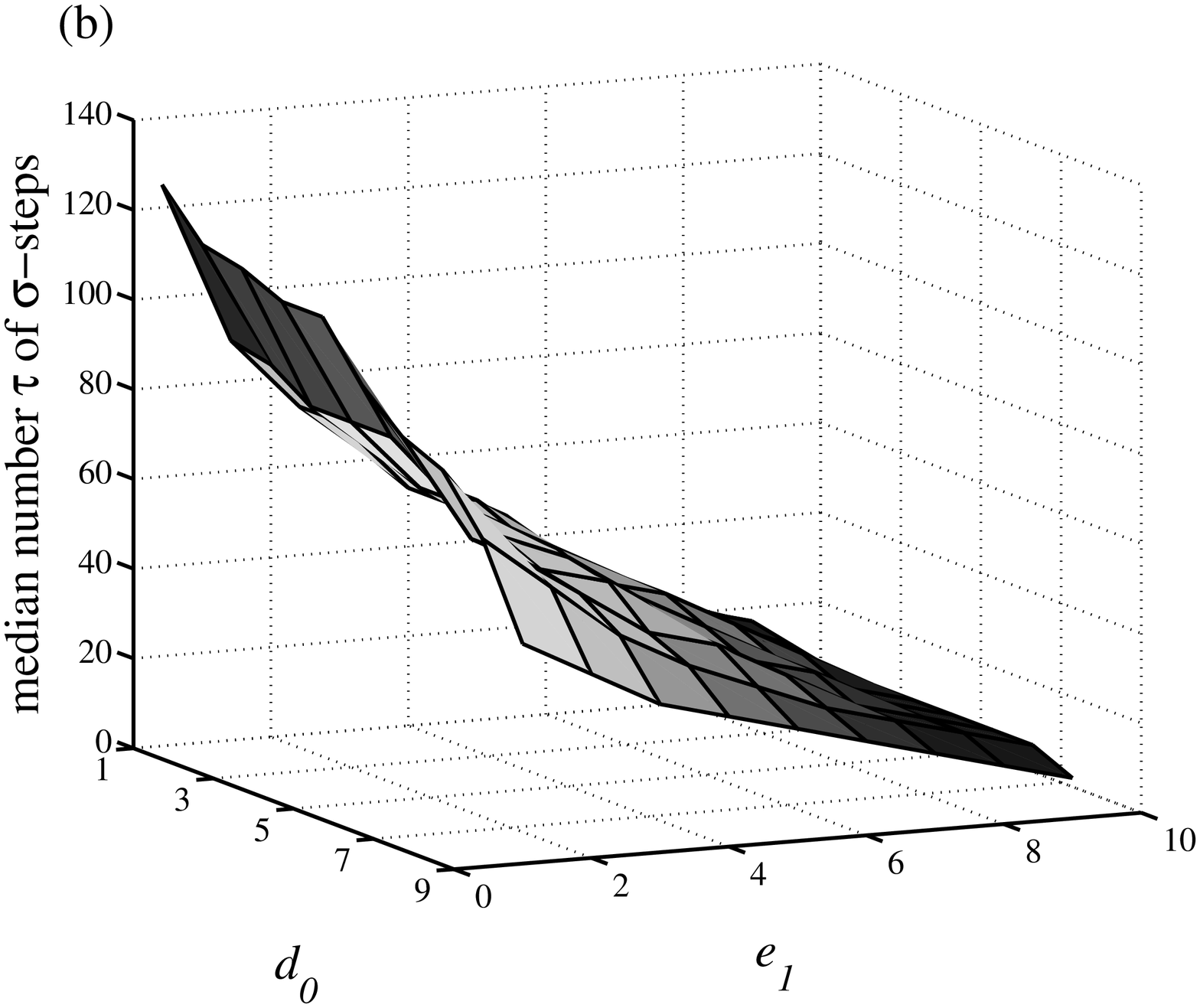}
\includegraphics[scale=0.32]{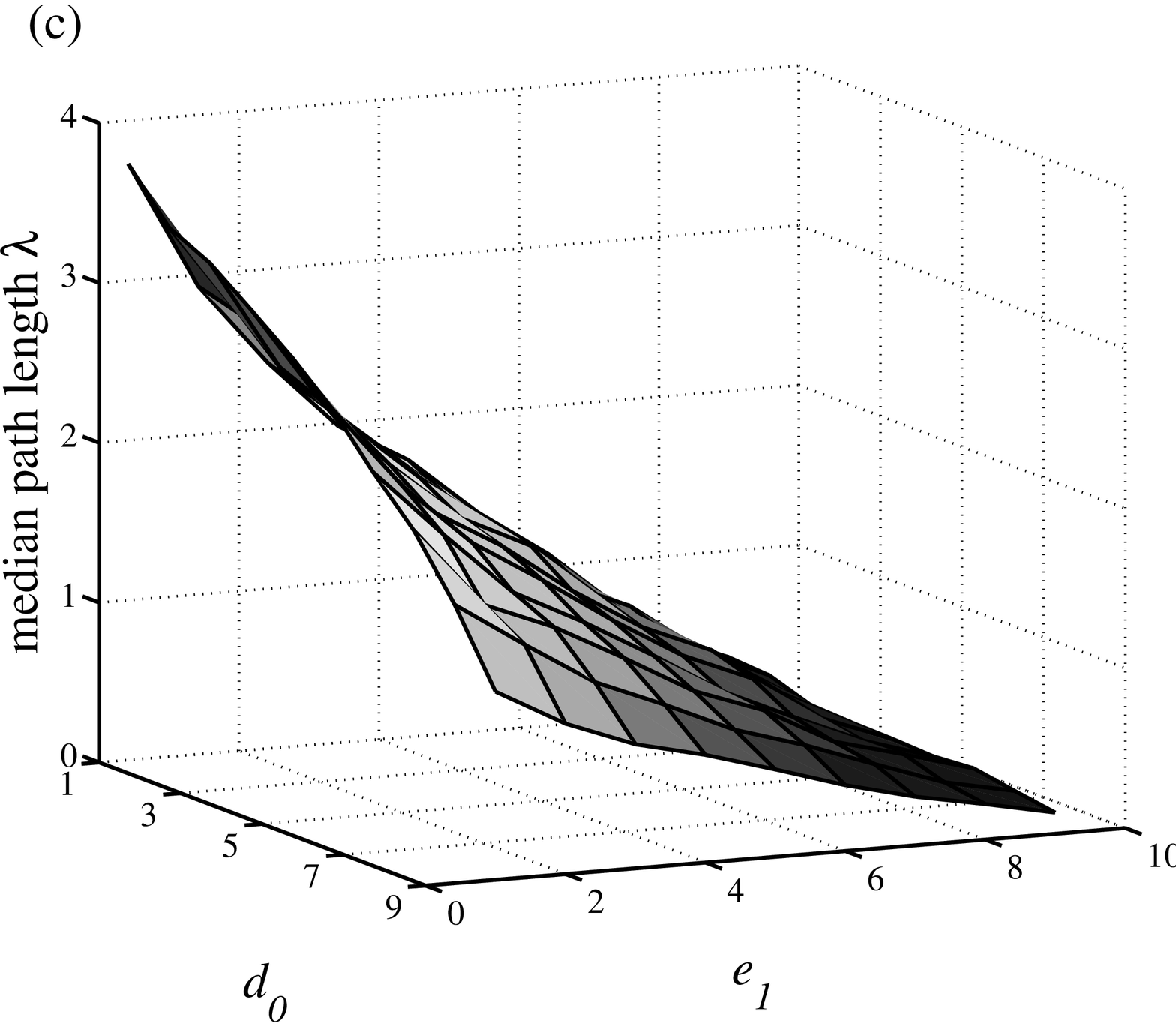}
\caption{Dependence of search efficiency on the degeneracy $d_0$ of the zero eigenvalue of $\rho$ and the degeneracy $e_1$ of the maximal eigenvalue of $\Theta$ for $N=10$. Figure (a) shows the dimension of the maximum submanifold as a function of these two parameters. Figure (b) shows the median number $\tau$ of $\sigma$-steps, and Figure (c) shows the median path length $\lambda$. Figures (b) and (c) show that the search efficiency increases as the dimension of the maximum submanifold increases.}
\label{3dplots1}
\end{figure}

In this section, we analyze the dependence of the optimization search effort on the degeneracy structure of $\rho$ and $\Theta$. Recall that ${\cal M}_{\rm max} = \{S \in {\cal S}: J(S) = \theta_{\rm max}\}$, where $\theta_{\rm max}$ is the maximal eigenvalue of $\Theta$. As shown in~\cite{wu1}, the dimension of ${\cal M}_{\rm max}$ is
\begin{equation}\label{dimMax}
\dim({\cal M}_{\rm max}) = 2(d_0+e_1)N^3-(2d_0e_1+1)N^2,
\end{equation}
where $d_0$ and $e_1$ are the degeneracies of the zero eigenvalue of $\rho$ and maximal eigenvalue $\theta_{\rm max}$ of $\Theta$, respectively. The dimension of the maximum manifold as a function of $d_0$ and $e_1$ is plotted on \Fref{3dplots1}~(a). If $d_0$ is close to $N$, then the initial state $\rho$ is close to a pure state, and for $e_1$ close to $N$, the target operator $\Theta$ is close to a constant multiple of the identity operator. Equation~(\ref{dimMax}) and \Fref{3dplots1}~(a) show that large values of $d_0$ and $e_1$ correspond to higher-dimensional maximum submanifolds (note that the dimension of the maximum manifold on vertical axis of \Fref{3dplots1}~(a) increases in the downward direction).

\begin{figure}\center
\includegraphics[scale=0.88]{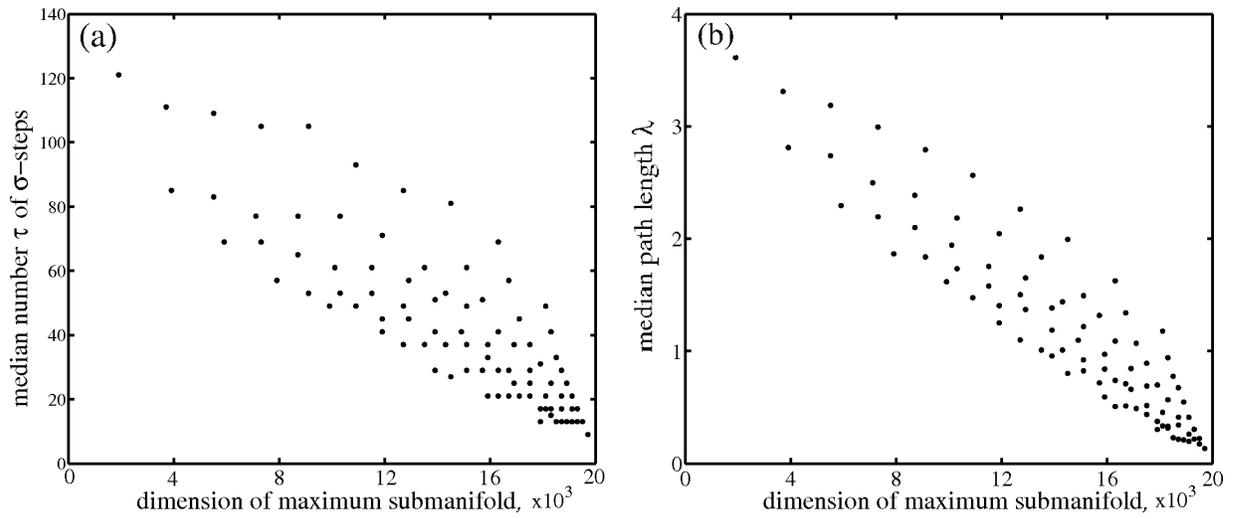}
\caption{Dependence of search efficiency on the dimension of the maximum submanifold for $N=10$. Figures (a) and (b) show the dependence of the median number $\tau$ of $\sigma$-steps and median path length $\lambda$, respectively, on the dimension of the maximum submanifold. The dimension of the maximum submanifold is determined by the pair $(d_0,e_1)$; each point on the plot corresponds to a $(d_0,e_1)$ pair. These figures show the increase in the search efficiency as the dimension of the maximum submanifold increases.}
\label{3dplots2}
\end{figure}

Figures~\ref{3dplots1}~(b) and~(c) show the dependence of efficiency of optimization upon $d_0$ and $e_1$. As $d_0$ and $e_1$ approach $N$, the efficiency of optimization increases rapidly. Comparison with \Fref{3dplots1}~(a) shows a strong positive correlation between the dimension of the maximum manifold and the efficiency of optimization. This result is expected, since an increase in $\dim({\cal M}_{\rm max})$ corresponds to a larger target submanifold of optimal solutions. The presence of the positive correlation is illustrated in a more explicit way in \Fref{3dplots2}, where the two parameters $\tau$ and $\lambda$ characterizing the efficiency of optimization are plotted versus the dimension of the maximum submanifold. The dimension of the maximum manifold is determined by the pair $(d_0,e_1)$ and different pairs can produce the same dimension of the maximum manifold. Each point in \Fref{3dplots2} corresponds to a pair $(d_0,e_1)$. The figure shows the general trend that an increase in the dimension of the maximum manifold decreases the required optimization search effort; however the correlation is not perfect and different pairs $(d_0,e_1)$ and $(d'_0,e'_1)$ with the same or almost the same dimensions of their respective maximum manifolds can have different values of the parameters $\tau$ and $\lambda$.

\subsection{Comparison of coherent and incoherent control}\label{unitcomp}
We now compare the efficiencies of coherent and incoherent control. The coherent control mechanism is implemented as follows. Let $S(0)$ be defined by $K_{i} = U/N$ for some $U \in {\cal U}(N)$. It is shown in \ref{invmanproof} that unitary Kraus maps form an invariant submanifold of ${\cal S}$ with respect to ${\rm grad }J(S)$. That is, if ${\cal S}_{\cal U} = \{S \in {\cal S}: \exists U \in {\cal U}(N): K_{i} = U/N\}$, then the solution to ${\rmd S}/{\rmd \sigma} = {\rm grad }J(S)$ with the initial condition $S(0) = S_0 \in {\cal S}_{\cal U}$ will lie entirely in ${\cal S}_{\cal U}$. Hence, solving the differential equation allows us to simulate density matrix evolution by coherent unitary control. Indeed, then $\rho_t = \sum_{i=1}^{N^2} K_{i}(t)\rho K_{i}^{\dagger}(t) = U(t)\rho U^{\dagger}(t)$.

\begin{figure}\center
\includegraphics[scale=0.37]{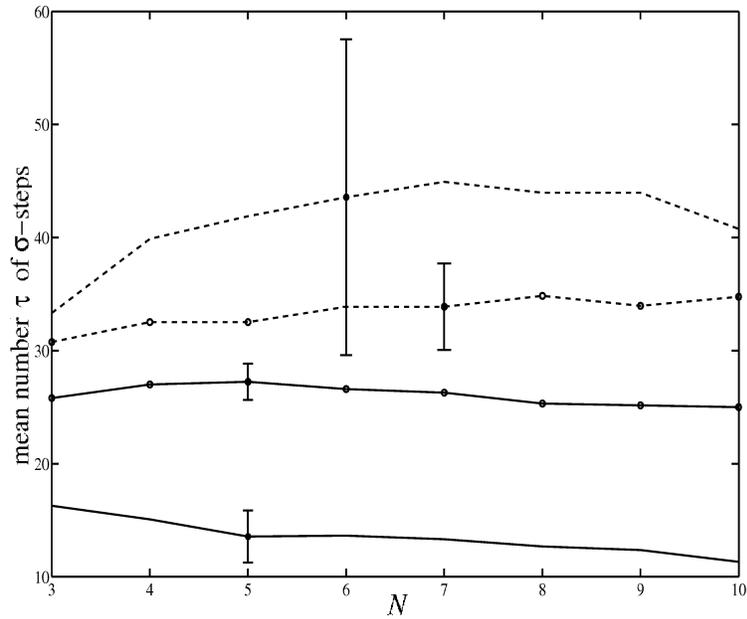}
\caption{Comparison of the optimization efficiency for incoherent control by Kraus maps and coherent control by unitary transformations, for both pure and mixed initial states $\rho$. Fifty simulations were performed to generate each point. The mean number $\tau$ of $\sigma$-steps with typical standard deviations indicated by error bars is plotted on the vertical axis. The solid lines correspond to control by Kraus maps, and the dashed lines correspond to control by unitary maps. The lines marked by circles correspond to pure state $\rho$, and the unmarked lines correspond to mixed $\rho$. Similar behavior is observed for path length (not shown).}
\label{unitarycomp}
\end{figure}

In all the simulations here, $\Theta = |N\rangle \langle N|$. For unitary control, the maximal value of the objective function $J[U] = \Tr[U \rho U^{\dagger}\Theta]$ is the maximal eigenvalue $\rho_{\rm max}$ of the initial state $\rho$. Thus, to ensure a fair comparison between the coherent and incoherent control, the target observable value is set to $J = \rho_{\rm max}$ for both incoherent and coherent control, and the algorithm stops as soon as the value $J=\rho_{\rm max}-0.01$ is attained. This stopping criteria is the reason for the difference between the curve corresponding to incoherent control in~\Fref{depN1} (a) and the curve in~\Fref{unitarycomp}; in the simulations displayed in the prior figure, the target observable value was $J=1$ rather than $J = \rho_{\rm max}$.

\Fref{unitarycomp} shows that with the ability to generate arbitrary Kraus maps, incoherent control can be a far more efficient process than coherent control for both pure and mixed $\rho$, especially for large values of $N$. The greater freedom allowed by incoherent control decreases the complexity of the problem and allows for a more efficient search.

\section{Control under linear constraints on the Kraus operators}\label{LinConst}
This section considers control under additional constraints on the available Kraus maps, which produce constraints on the Stiefel manifold. The target operator is assumed to have the form $\Theta = |N\rangle\langle N|$.

Let $h:{\cal S} \rightarrow \mathbb{R}^q$ be a set of $q$ real-valued constraints. Recall from \sref{stiefform1} that the objective function $J$ is invariant under ${\cal W}$-transformations. Since ${\cal W}$-transformations correspond to different parametrizations of the same physical evolution Kraus map, any reasonable constraint should be ${\cal W}$-invariant, and thus we impose the requirement that $h(S) = h(WS)$ for any $S \in {\cal{S}}$ and any $W \in {\cal{W}}$.

We restrict the attention to {\it affine} constraints, which are of the form $h(S) = g(S)-\gamma$, where $g$ is linear over $\mathbb{R}$ and $\gamma \in \mathbb{R}^q$ is a constant. Specifically, for a given set of matrices $\{B_1,\dots,B_n\}$ we consider ${\cal W}$-invariant affine constraints of the form ${\rm Tr}(B_l^{\dagger}K_j) = 0$ for each $l=1,\dots,n$, $j=1,\dots,N^2$, and $B_l \in {\cal M}_N$. Since ${\rm Tr}(B^{\dagger}_l\tilde{K_j})={\rm Tr}\left(\sum_{i=1}^{N^2} u_{ji}B^{\dagger}_lK_i\right) = 0$ by linearity of the trace operation, this constraint is ${\cal W}$-invariant and the set of Kraus matrices satisfying this constraint forms a ${\cal W}$-invariant subset of the Stiefel manifold.

\subsection{Numerical procedure}
The constraints ${\rm Tr}(B^{\dagger}_iK_j) = 0$ with $1 \leq i \leq n$, $1 \leq j \leq N^2$ can be rewritten as a set of $2nN^2$ constraints $h_k: {\cal S}\rightarrow \mathbb{R}$ defined as follows. Let $\tilde G_l \in {\cal M}(N^3,N,\mathbb{C})$ for $l=n(j-1)+i$ be the matrix with $B_i$ occupying rows $(j-1)N+1$ through $jN$ and with other matrix elements set to zero. Then define $G_k=\tilde G_k$ for $k=1,\dots,nN^2$, $G_k={\rm i}\tilde G_{k-nN^2}$ for $k=nN^2+1,\dots,2nN^2$ and set $h_k(S) =\langle G_k,S\rangle$. The control goal is to maximize $J$ over $h^{-1}(0)$.

First, we need to find a matrix $\tilde{S} \in {\cal S}_h$ which represents an initial control satisfying the constraint. To do this, define
\begin{equation}
f(S) = \sum_{k=1}^{2nN^2} \langle G_k,S\rangle^2.
\end{equation}
We see that
\begin{eqnarray}
{\rmd}_Sf(\delta S) &= \sum_{k=1}^{2nN^2} 2\langle G_k,\delta S\rangle \langle G_k,S\rangle =\Bigl\langle \sum_{k=1}^{2nN^2}2\langle G_k,S\rangle G_k,\delta S\Bigr\rangle.
\end{eqnarray}
Hence, ${\rm grad }\, f(S) = \sum_{k=1}^{2nN^2}2\langle G_k,S\rangle G_k$. Now generate an arbitrary $S_0 \in {\cal S}$ and solve the equation
\[
\diff{S}{\sigma} = -{\cal P}_{S}({\rm grad }\, f(S)),
\]
with the initial condition $S(0) = S_0$, where ${\cal P}_S$ is the orthogonal projector from ${\cal M}(N^3,N,\mathbb{C})$ onto $T_S{\cal S}$ (see Section~\ref{alg}). Then, if the landscape of $f$ on ${\cal S}$ is trap-free, the algorithm will always find a global minimum $\tilde{S}$ of $f$, which will satisfy the constraint $h(\tilde{S}) = 0$. It is unknown whether this constrained landscape is trap free.

After producing the initial Stiefel matrix $\tilde S$, we maximize the objective function $J$ on ${\cal S}_h$ by solving the differential equation
\[
\diff{S}{\sigma} = {\cal P}_{h,S}({\rm grad }\, J(S)),
\]
with the initial condition $S(0) = \tilde{S}$. Here ${\rm grad }\, J(S)$ is the gradient of $J$ on ${\cal S}$ and ${\cal P}_{h,S}$ is a projector from $T_S{\cal S}$ onto $T_S{\cal S}_h$. The expicit expression for ${\cal P}_{h,S}$ is derived in~\ref{appendixB}.

\subsection{Numerical results: general linear constraint}
It is difficult to derive a general analytical expression for the maximum value of $J$ on the constrained manifold $h^{-1}(0)$ due to the complicated nature of the constraints. For this reason, we cannot determine that the gradient algorithm is stuck at a false trap $\tilde{S}$ (where ${\rm grad}J(\tilde{S}) = 0$) by simply calculating $J(\tilde{S})$. Therefore, for a fixed constraint $h$, we performed the optimization procedure ten times using a different initial condition $\tilde S(0)$ each time and compared the resultant ten maximal values of the objective function. Let $S^*_i$ be the optimal control on the $i^{\rm th}$ run (where $1 \leq i \leq 10$), with corresponding maximum value $J^*_i = J(S^*_i)$. If $J^*_k < J^*_l$ for some $k$ and $l$, then $S^*_k$ is a false trap. Note that $J^*_k = J^*_l$ for all $k$ and $l$ does not guarantee that the landscape is trap-free; the only conclusion is that the ten runs of the algorithm have not found a false trap.

We performed simulations for $N=2, 3, 4$. For each $N$, five different initial states $\rho$ were generated, and for each $\rho$ five different collections of matrices $\{B_1,\dots,B_n\}$ corresponding to five constraints were produced. We consider $n \leq N^2-N-1$, where $N^2-N-1$ represents the maximum number of constraints of special form corresponding to fixing to zero individual matrix elements of the Kraus operators. As a result of the numerical optimization, each of the ten runs performed with initial controls $\tilde S_i(0)$ produced the same maximal value $J^*=J^*_i$, and therefore we did not find a false trap. Although this result does not prove the absence of false traps for linear constraints, it indicates that it is surprisingly difficult to find such traps, if they exist.

\subsection{Numerical results: fixing to zero individual matrix elements}\label{fixelem}
We now consider a special case of the ${\cal W}$-invariant linear constraints such that $h: {\cal S}\rightarrow \mathbb{R}^{2N^2}$ is the constraint $(K_l)_{ij} = 0$ for all $l=1,\dots,N^2$ and for some pair $(i,j)$. The constraint corresponds to setting the ${ji}^{\rm th}$ element in each of the $N^2$ Kraus operators to zero; we consider the real and imaginary parts separately, hence there are $2N^2$ constraints. Since $\tilde{K}_n = \sum_{m=1}^{N^2} u_{nm}K_{m}$ defines the ${\cal W}$-transformation, $(\tilde{K}_n)_{ij} = 0$ for all $n$ as well. Hence, $h(WS) = h(S)$, and the constraint is ${\cal W}$-invariant. More generally, we consider ${\cal W}$-invariant constraints of the form
\begin{equation}
 (K_l)_{i_q,j_q}\equiv (Y_{j_q}^{i_q})_l = 0, \qquad l=1,\dots,N^2,\quad \forall j_q \in I_1,\quad \forall i_q \in I_2,\quad q=1,\dots,n,
\end{equation}
where $I_1$ and $I_2$ two subsets of the set $\{1,2,\dots,N\}$ each with $n$ elements.

For such a constraint, equation~(\ref{Yform}) can be used to determine analytically the optimal value of the objective function $J$ on the constrained set $h^{-1}(0)$:
$$
J_{\rm max} = \left\{ % use packages: array
\begin{array}{l}
1 \qquad\qquad\qquad\, \textrm{ if } N \notin I_1 \\
1-\sum_{j \in I_2} \rho_{jj} \quad \textrm{ if } N \in I_1
\end{array}\right.
$$
For each $N=2,3,4,5$, we fix to zero $n$ matrix elements of every Kraus operator, with $n$ between $N$ and $N^2-N-1$ (the maximum possible number of matrix elements which can simultaneously be fixed to zero). For a given $n$, the optimization procedure was performed 25 times, and a different collection of matrix elements was fixed to zero during each run (i.e., different sets $I_1$ and $I_2$ were choosen). The gradient algorithm was able to reach the maximal value $J_{\rm max}$ each time, showing that there do not appear to be false traps in this landscape. If suboptimal maxima were encountered, the algorithm would have gotten stuck at $J < J_{\rm max}$, and global optimization could not have been performed. Thus the optimization procedure did not discover any false traps for $25$ randomly generated constraints. Again, this could not be taken as conclusive proof of the absence of false traps. Evidently, more complex or demanding constraints are called for to find traps.

\section{Conclusion}\label{conclusions}
This paper analyzes the efficiency of optimization over control landscapes for open quantum systems governed by Kraus map evolution. Several conclusions stem from the findings. When $\Theta$ is a rank-one projector, which corresponds to the control goal of transforming an initial state $\rho$ into a pure state, the search efficiency primarily depends on the number of nonzero eigenvalues of the initial state. The efficiency is relatively insensitive to the dimension of the quantum system $N$, provided that the number of populated energy states in the initial density matrix remains constant. As the number of nonzero eigenvalues of $\rho$ rises with $N$, the search for an optimal control becomes less efficient. This result agrees with the expectation that transforming a high-entropy initial state into a low-entropy final state is a more difficult control problem than controlled transformations between states with similar entropy.

The analysis also reveals that for fixed $N$, the search efficiency positively correlates with the number of zero eigenvalues of $\rho$. This result can be extended to a more general principle: when the dimension of the quantum system is fixed, the dimension of the maximum submanifold (the set of Kraus operators that correspond to optimal control) positively correlates with the efficiency of the optimization procedure. This statement agrees with the common intuition that a \lq \lq larger\rq \rq target results in an easier and more efficient search. The scaling behavior with $N$ found in this work is also consistent with that identified with unitary evolution, both dynamically and kinematically~\cite{moore08,riviello08}.

We then showed that incoherent control modelled by Kraus map evolution, under the assumption that any Kraus map can be generated, is more efficient than coherent control modelled by unitary evolution. The larger number of control variables available in incoherent control actually decreases the complexity of the search effort. While the influence of the environment makes the total system ostensibly more complicated, the results show that the ability to control the environment can decrease the search effort.

We also analyzed control landscapes with linear constraints on the Kraus maps. Even with the maximum possible number of linear constraints, false traps were not found. While this result does not prove the absence of false traps, it is nonetheless surprising. In the future work, we would like to investigate the control landscapes for constrained Kraus maps in more detail both numerically and theoretically.

The kinematic analysis needs to be extended by a more detailed investigation of the role of the critical structure of the control landscapes on the search effort. In particular, the possible influence of saddle manifolds on the required search effort should be analyzed. This analysis may reveal more subtle structural details about the quantum control landscapes. Also non-topological properties of quantum control landscapes may affect the optimization efficiency. In general, it is necessary to find all essential characteristics of the initial state $\rho$ and the target operator $\Theta$ that affect the efficiency of the search. Another important problem is to study the dynamics of controlled open quantum systems with regard to topological and non-topological characteristics of the corresponding dynamical control landscapes. Various specific model systems can be used to study the dependence of search efficiency upon the parameters characterizing the system and environment. The presence or absence of false traps in the dynamical control landscapes should be investigated, including situations with constraints on the dynamical controls.

\section{Acknowledgements}
The authors acknowledge support from the NSF and ARO. A. Pechen also
acknowledges partial support from the grant RFFI 08-01-00727-a.

\appendix
\section{Appendix A. Invariance of the submanifold ${\cal S}_{\cal U}$ for $\rmd S/\rmd \sigma = {\rm grad}\, J(S(\sigma))$}\label{invmanproof}
Here we show that the submanifold ${\cal S}_{\cal U} := \{S_U \in {\cal S}\, |\, \exists U \in {\cal U}(N) \textnormal{ such that } K_i = \frac{1}{N}U\textnormal{ for } i=1,\dots,N^2\}$ (i.e., all of the Kraus matrices determining a point $S_U \in {\cal S}_{\cal U}$ are equal to the same constant multiple of some unitary matrix) is invariant for the differential equation $\rmd S / \rmd \sigma = {\rm grad }\, J(S(\sigma))$.

Let $X$ be a manifold with tangent bundle $TX$. Consider the differential equation
\begin{equation}\label{diffeq}
\rmd x/\rmd \sigma = f(x(\sigma))
\end{equation}
where $f: X \rightarrow TX$ is a smooth function, and $x: [0,1] \rightarrow X$ is a path through $X$ parametrized by the real variable $\sigma$. A manifold $Y \subset X$ is called an {\it invariant submanifold} for the differential equation (\ref{diffeq}) if $x(0) \in Y$ implies that $x(\sigma) \in Y$ for all $\sigma \in [0,1]$. A compact manifold $Y\subset X$ is an invariant submanifold for (\ref{diffeq}) if and only if $f(x) \in T_x Y$ for each $x \in Y$~\cite{chicone1}.

It was shown in \sref{alg} that $\delta S \in T_S{\cal S}$ if and only if $S^{\dagger}(\delta S)$ is skew-Hermitian. Therefore, writing $\delta S$ as a stack of $N^2$ $N \times N$ matrices $\delta S_1,\dots,\delta S_{N^2}$, we see that for any $S_U \in {\cal S}_{\cal U}$, $\delta S \in T_{S_U}{\cal S}_{\cal U}$ if the matrix $U^{\dagger}\sum_{k=1}^{N^2}\delta S_{k}$ is skew-Hermitian.

\begin{theorem} Let $\be$ be the $N^2 \times 1$ matrix (column vector) with all elements equal to one (i.e., $\be(i) = 1$ for all $i$). Then for any $S = \frac{1}{N}(\be \otimes U) \in {\cal S}_{\cal U}$, the matrix $Z:=U^{\dagger}[(\be^{\dagger} \otimes {\mathbb I}_N)\ {\rm grad}\, J(S)]$ is skew-Hermitian.
\end{theorem}
{\bf Proof.} Recall that ${\rm grad}\, J(S) = (2{\mathbb I}_{N^3}-SS^{\dagger})({\mathbb I}_{N^2} \otimes \Theta)S\rho - S\rho S^{\dagger}({\mathbb I}_{N^2} \otimes \Theta)S$. Then
\begin{eqnarray}
 Z&=&U^{\dagger}(\be^{\dagger} \otimes {\mathbb I}_N)\Bigl[\frac{2}{N}{\mathbb I}_{N^3} -\frac{1}{N^3}(\be \otimes U)(\be^{\dagger} \otimes U^{\dagger})\Bigr]({\mathbb I}_{N^2} \otimes \Theta)(\be \otimes U)\rho  \nonumber \\
&&- U^{\dagger}\frac{1}{N^3}(\be \otimes U)\rho (\be^{\dagger} \otimes U^{\dagger})({\mathbb I}_{N^2} \otimes \Theta)(\be \otimes U) \nonumber \\
&=& U^{\dagger}\Bigl[\frac{2}{N}(\be^{\dagger} \otimes {\mathbb I}_N){\mathbb I}_{N^3}-\frac{1}{N} U(\be^{\dagger} \otimes U^{\dagger})\Bigr]({\mathbb I}_{N^2} \otimes \Theta)(\be \otimes U)\rho \nonumber \\
&&- U^{\dagger}\frac{1}{N} U\rho (\be^{\dagger} \otimes U^{\dagger})({\mathbb I}_{N^2} \otimes \Theta)(\be \otimes U) \nonumber \\
&=& \Bigl[\frac{2}{N}U^{\dagger}(\be^{\dagger} \otimes {\mathbb I}_N){\mathbb I}_{N^3}-\frac{1}{N}(\be^{\dagger} \otimes U^{\dagger})\Bigr]({\mathbb I}_{N^2} \otimes \Theta)(\be \otimes U)\rho \nonumber \\
&&-\frac{1}{N}\rho(\be^{\dagger} \otimes U^{\dagger})({\mathbb I}_{N^2} \otimes \Theta)(\be \otimes U) \nonumber \\
&=& \Bigl[\frac{2}{N}U^{\dagger}(\be^{\dagger} \otimes {\mathbb I}_N){\mathbb I}_{N^3}-\frac{1}{N}(\be^{\dagger} \otimes U^{\dagger})\Bigr](\be \otimes \Theta U)\rho - N\rho U^{\dagger}\Theta U \nonumber \\
&=& 2N U^{\dagger}\Theta U\rho - N U^{\dagger}\Theta U\rho - N\rho U^{\dagger}\Theta U \nonumber \\
&=& N[U^{\dagger}\Theta U,\rho]
\end{eqnarray}
which is skew-Hermitian for Hermitian matrices $\rho$ and $\Theta$. As a result, ${\rm grad }\, J(S_U) \in T_{S_U}{\cal S}_{\cal U}$ for $S_U \in {\cal S}_{\cal U}$, so ${\cal S}_{\cal U}$ is an invariant submanifold for the differential equation $\rmd S / \rmd \sigma = {\rm grad }\, J(S(\sigma))$.

\section{Appendix B. Derivation of the projector ${\cal P}_{h,S}$}\label{appendixB}
Let $\tilde{h}: {\cal M}(N^3,N,\mathbb{C}) \rightarrow \mathbb{R}^q$ define a constraint on the Stiefel matrices, which restricts the set of addmissible controls to ${\cal S}_h = {\cal S} \cap \tilde{h}^{-1}(0)$. The goal is to find a projector ${\cal P}_{h,S}: T_S{\cal S} \rightarrow T_S{\cal S}_h$, such that the gradient of $J$ on ${\cal S}_h$ will be ${\cal P}_{h,S}({\rm grad }\, J(S))$.

We will use the following lemma.
\begin{lemma}\label{lemma_appendixB}
Let $X$ and $Y$ be Riemannian manifolds and $F: X \rightarrow Y$. Suppose that $\rmd_x F$ is surjective for all $x \in X$. Let $P_x$ be the operator on $T_xX$ defined as $P_x = I - (\rmd_xF)^* \circ (\rmd_x F \circ (\rmd_x F)^*)^{-1} \circ \rmd_xF$. Then {\bf (a)} $P_x$ is a projection (that is, ${P_x}^2 = P_x$) and \textbf{(b)} $P_x: T_xX \rightarrow T_x F^{-1}(F(x))$.
\end{lemma}
{\bf Proof.} \textbf{(a)}. It is straightforward to see that $P_x^2 = P_x$:
\begin{eqnarray}
P_x^2 &=& (I - (\rmd_xF)^* \circ (\rmd_x F \circ (\rmd_x F)^*)^{-1} \circ \rmd_xF)^2 \nonumber \\ &=& I - 2(\rmd_xF)^* \circ (\rmd_x F \circ (\rmd_x F)^*)^{-1} \circ \rmd_xF \nonumber \\
&& + (\rmd_xF)^* \circ (\rmd_x F \circ (\rmd_x F)^*)^{-1} \circ \rmd_xF \circ (\rmd_xF)^* \circ (\rmd_x F \circ (\rmd_x F)^*)^{-1} \circ \rmd_xF \nonumber \\
& =& I - 2(\rmd_xF)^* \circ (\rmd_x F \circ (\rmd_x F)^*)^{-1} \circ \rmd_xF
+ (\rmd_xF)^* \circ (\rmd_x F \circ (\rmd_x F)^*)^{-1} \circ \rmd_xF \nonumber \\
& =& I - (\rmd_xF)^* \circ (\rmd_x F \circ (\rmd_x F)^*)^{-1} \circ \rmd_xF=P_x
\end{eqnarray}
\textbf{(b)}. It is clear that if $\rmd_xF(z) = 0$, then $z \in T_xF^{-1}(F(x))$. Note that any vector $v \in T_xX$ can be written as $v = z+\rmd_x F^*(w)$, where $w$ is arbitrary and $z \in T_xF^{-1}(F(x))$. Indeed, let $w = (\rmd_x F \circ (\rmd_xF)^*)^{-1} \circ \rmd_xF(v)$. Then $\rmd_xF(z) = \rmd_xF(v)-\rmd_xF \circ (\rmd_x F)^* \circ (\rmd_x F \circ (\rmd_xF)^*)^{-1} \circ \rmd_xF(v)=0$, and therefore $z \in T_xF^{-1}(F(x))$.

Now we will show that the image of $P_x$ lies in $T_xF^{-1}(F(x))$. For any $v \in T_xX$, write $v = z+\rmd_xF^*(w)$, where $z \in T_xF^{-1}(F(x))$. Then $P_x(v) = P_x(z) + P_x \circ (\rmd_xF)^*(w) = z - (\rmd_xF)^* \circ (\rmd_x F \circ (\rmd_x F)^*)^{-1} \circ \rmd_xF(z) + (\rmd_xF)^*(w) - (\rmd_xF)^* \circ (\rmd_x F \circ (\rmd_x F)^*)^{-1} \circ \rmd_xF \circ (\rmd_xF)^*(w) = z - (\rmd_xF)^* \circ (\rmd_x F \circ (\rmd_x F)^*)^{-1} \circ \rmd_xF(z) = z$ since $z \in T_xF^{-1}(F(x))$ by assumption. Hence, the image of $P_x$ lies in $T_xF^{-1}(F(x))$. This proves the lemma.

Recall now that ${\cal P}_S$ is the projector from ${\cal M}(N^3,N,\mathbb{C})$ to $T_S{\cal S}$. Then ${\rmd}_Sh = {\rmd}_S\tilde{h}|_{T_S{\cal S}}$ and ${\rmd}_Sh^* = {\cal P}_S \circ {\rmd}_S\tilde{h}^*$. If ${\rmd}_Sh$ is full-rank, then according to lemma~\ref{lemma_appendixB} we have a projector ${\cal P}_{h,S}: T_S{\cal S} \rightarrow T_S{\cal S}_h$:
\begin{eqnarray}
{\cal P}_{h,S}(\delta S) &=& \delta S - {\rmd}_S h^* \circ ({\rmd}_S h \circ {\rmd}_S h^*)^{-1} \circ {\rmd}_S h(\delta S) \nonumber \\
&=& \delta S - {\cal P}_S \circ {\rmd}_S\tilde{h}^*\circ ({\rmd}_S\tilde{h}\circ {\cal P}_S \circ {\rmd}_S\tilde{h}^*)^{-1} \circ {\rmd}_S\tilde{h}(\delta S). \nonumber
\end{eqnarray}

In what follows, we will restrict our attention to affine maps defined by a set $\tilde{g} = (\tilde{g}_1,\dots,\tilde{g}_q)$ of bounded linear functionals $\tilde{g}_i: {\cal M}(N^3,N,\mathbb{C})\rightarrow \mathbb{R}$. By the Riesz Representation Theorem, there exist unique matrices $G_i \in {\cal M}(N^3,N,\mathbb{C})$ such that $\tilde{g}_i(A) = \langle G_i,A\rangle$ for all $A \in {\cal M}(N^3,N,\mathbb{C})$. For a constraint of the form $h_k(S) = \Re(Y_j^{i})_k = 0$ (resp. $\Im(Y_j^{l})_k = 0$) as considered in \sref{fixelem}, these matrices have the form $G_k = |d\rangle\langle l|$ (resp. $G_k = i|d\rangle\langle l|$), where $d=(k-1)N^2+(j-1)N$.

Since $\gamma$ is constant and $\tilde{g}$ is linear, ${\rmd}_S\tilde{h} = \tilde{g}$ and ${\rmd}_S\tilde{h}^* = \tilde{g}^*$. To determine a formula for $\tilde{g}^*$, note that for any $y \in \mathbb{R}^q$
\[
\langle \tilde{g}^*(y),\delta S\rangle = \langle y, \tilde{g}(\delta S)\rangle = \sum_{i=1}^q y_i\tilde{g}_i(\delta S) = \sum_{i=1}^q y_i \langle G_i,\delta S\rangle = \biggl\langle \sum_{i=1}^q y_iG_i,\delta S\biggr\rangle.
\]
Therefore, $\tilde{g}^*(y) = \sum_{i=1}^q y_iG_i$. Putting these expressions together gives
\begin{eqnarray}
{\rmd}_S\tilde{h} \, \circ \, {\rmd}_S \tilde{h}^*(y) &=& \tilde{g}\left(\sum_{i=1}^q y_i{\cal P}_S(G_i)\right) = \pmatrix{\langle G_1,\sum_{i=1}^q y_i{\cal P}_S(G_i)\rangle \cr \vdots \cr \langle G_q,\sum_{i=1}^q y_i{\cal P}_S(G_i)\rangle} \nonumber\\
&=& \sum_{i=1}^q y_i\pmatrix{\langle G_1,{\cal P}_S(G_i)\rangle \cr \vdots \cr \langle G_q,{\cal P}_S(G_i)\rangle} = Zy, \nonumber
\end{eqnarray}
where $Z$ has matrix elements $Z_{ij} = \langle G_i,{\cal P}_S(G_j)\rangle$. We finally get
\begin{eqnarray}
 {\cal P}_{h,S}(\delta S) &=& \delta S - {\cal P}_S \circ \tilde{g}^*(Z^{-1}\tilde{g}(\delta S))\nonumber \\
&=& \delta S - \sum_{i=1}^q  {\cal P}_S(G_i)(Z^{-1}\tilde{g}(\delta S))_i \nonumber \\
&=& \delta S - \sum_{i,j=1}^q  {\cal P}_S(G_i)(Z^{-1})_{ij}\tilde{g}_j(\delta S)\nonumber \\
&=& \delta S - \sum_{i,j=1}^q {\cal P}_S(G_i)(Z^{-1})_{ij}\langle G_i,\delta S\rangle. \nonumber
\end{eqnarray}

\end{document}